\documentclass[twocolumn]{aastex631}
\usepackage[utf8]{inputenc}
\usepackage{color}
\usepackage{graphicx}	
\usepackage{amsmath}	
\usepackage{wasysym}
\usepackage{mathtools}
\usepackage{newtxtext,newtxmath}
\usepackage{appendix}
\usepackage{longtable}

\newcommand{\hi}{{\rm H}{\textsc i}~}
\newcommand{\him}{{\rm H}{\textsc i}}
\newcommand{\dSFRMS}{$\rm \Delta SFR_{MS}$}

\shorttitle{Correlation between morphology and SF}
\shortauthors{Zhou et al.}

\begin{document}
\title{The link between galaxy structure properties and star formation in local galaxies}

\author[0000-0002-4135-0977]{Zhimin Zhou}
\affil{National Astronomical Observatories, Chinese Academy of Sciences, Beijing 100101, People’s Republic of China}
\correspondingauthor{Zhimin Zhou}
\email{zmzhou@nao.cas.cn}

\author{Wenwen Wang}
\affil{National Astronomical Observatories, Chinese Academy of Sciences, Beijing 100101, People’s Republic of China}
\affil{College of Astronomy and Space Sciences, University of Chinese Academy of Sciences, Beijing 100049, People’s Republic of China}

\begin{abstract}
To investigate the role of morphology in galaxy evolution, we analyze the relationships between galaxy structure, star formation, and \hi gas content. Using multi-band images from the DESI Legacy Imaging Surveys, we perform detailed structural decompositions on a representative local galaxy sample from xGASS. Structural components and color properties are examined as functions of deviations from the star formation main sequence (\dSFRMS) and \hi gas deficiency ($\rm \Delta f_{\him}$).
We find that bulge fractions decrease with higher \dSFRMS\ and lower stellar mass, indicating that star-forming galaxies are predominantly disc-dominated, while quiescent galaxies are bulge-dominated. The slope of the color ($g-r$) versus \dSFRMS\ relationship decreases from low to high stellar masses and from outer to inner regions, with greater color variation in massive galaxies. Color gradients are predominantly negative, becoming shallower in lower-mass galaxies and in the outer disk regions.
We also identify inflection points in the color gradient and bulge fraction relations with \dSFRMS, with main-sequence galaxies having the lowest bulge fractions and steepest color gradients. At fixed stellar mass, we observe only a slight correlation between bulge fraction and \hi deficiency. However, outer disk colors show a stronger dependence on \hi content than inner regions, and color gradients flatten as $\rm \Delta f_{\him}$ increases.
These results suggest that \hi gas is more closely linked to star-forming, disc-dominated systems, supporting the idea that gas accretion fuels star formation primarily in galaxy disks.
\end{abstract}


\section{Introduction} \label{sec:intro}
Galaxies in the local universe have long been known to exhibit strong bimodality in color-magnitude and star formation rate (SFR)-stellar mass distributions, dividing them into either the blue cloud or red sequence, or equivalently, into star-forming and quiescent categories \citep[e.g.,][]{Blanton2003, Kauffmann2003, Brammer2009, Taylor2015}. Star-forming galaxies are generally blue, gas-rich, and disk-dominated, with significant ongoing star formation, while quiescent galaxies are typically massive, gas-poor, and bulge-dominated, displaying red colors and older stellar populations \citep{Bell2003, Schawinski2014, Barro2017, Leroy2019}.

Understanding how galaxies cease star formation and transform morphologically from disks to spheroids has long been a challenge. Numerical studies have investigated the physical processes and mechanisms behind quenching. For instance, by examining the relationships between galaxy mass, SFR, and environment, \citet{Peng2010} demonstrated that two distinct processes, known as mass quenching and environment quenching, may drive the transition from star-forming to passive galaxies \citep{Baldry2004, Kauffmann2004, Tomczak2014, Woo2015}.

As a fundamental characteristic of galaxies, morphology plays a key role in quenching and transformation processes and is essential for describing galaxy properties and understanding their formation, evolution, and dynamics \citep{Snyder2015}. The relationships between galaxy structures (such as the bulge, disk, and bar) and star formation activity offer valuable insights into the evolutionary history of galaxies \citep{Wuyts2011, Hopkins2013}.

For instance, the development of bulge and disk components in galaxies can help constrain quenching mechanisms based on observational data \citep[e.g.,][]{Fang2013, Gillman2024}. Bulges, characterized by a high S\'{e}rsic index (typically $\it n = 4$), are thought to form through galaxy mergers \citep{Bell2012} or other violent processes \citep{Inoue2012}. Beyond these classical de Vaucouleurs-like bulges, pseudo-bulges, commonly found in late-type spirals, exhibit lower S\'{e}rsic indices and disk-like features, and are likely products of secular evolution in galaxies \citep{Kormendy2004, Fisher2008, Hu2024}. Additionally, star formation can be quenched when gas is either removed or stabilized within a galaxy, potentially driven by bulge-related processes such as active galactic nucleus feedback or morphological quenching \citep{Bell2008, Fang2013}.

In contrast, galaxy disks typically exhibit an exponential light profile, corresponding to a S\'{e}rsic index of $\it n = 1$ \citep{Blanton2003}. Disk-dominated galaxies are generally bluer and exhibit higher star formation activity than bulge-dominated galaxies \citep{Snyder2015}, though exceptions exist, such as red disk galaxies with minimal or no star formation \citep{Hao2019, Wangl2022, Li2024} and blue, bulge-dominated galaxies with intense star formation \citep{Kannappan2009, Davis2011, Biswas2024}. The development of disk structures is often associated with ongoing star formation, gas accretion, and accumulation, suggesting different formation processes compared to bulges \citep{van2011, Chen2020}.

Structural decomposition of galaxies is a powerful tool for investigating and understanding their evolution. In recent years, a variety of automated software suites have been developed for characterizing bulge and disk properties through 2D decomposition of galaxy images, including GALFIT \citep{Peng2002, Peng2010}, GIM2D \citep{Simard2002}, BUDDA \citep{de2004, Gadotti2008}, Galapagos \citep{Barden2012}, Profit \citep{Robotham2017}, and SExtractor++ \citep{Bertin2020, Kummel2020}. These tools have enabled the creation of large galaxy samples to reveal stellar structures in both local and high-redshift galaxies \citep[e.g.,][]{Lackner2012, Lima-Dias2024, Nedkova2024}. For instance, significant efforts have been made to decompose the images of millions of galaxies from the Sloan Digital Sky Survey \citep[SDSS;][]{Simard2011, Meert2015}.

However, several factors must be considered to enhance the reliability and accuracy of statistical studies in structural decomposition. In addition to flexible and automated algorithms, these decompositions require large datasets with sufficiently deep imaging and high spatial resolution \citep{Euclid2023}. Another key factor is the quantity and types of fitting components included in the decompositions. While most recent studies have conducted bulge+disk decompositions on large galaxy samples, omitting other structures, such as bars and rings, can lead to overestimating the bulge contribution when these features reside in galaxy centers \citep{Nedkova2024}. For example, \citet{Morselli2017} compared bulge-disk decomposition results using the same image datasets and found that excluding bars led to a 10\%–25\% overestimation in the bulge-to-total light ratio (B/T), significantly steepening the slope gradient in related relationships \citep{Gadotti2008, Meert2015}.

To understand how galaxy morphology correlates with star formation and gas content, precise estimation of structural properties is essential. In this paper, we present a detailed analysis of structural decompositions for a representative galaxy sample, xGASS, using a homogeneous dataset of optical images from the DESI Legacy Imaging Surveys. These data allow us to explore structural and stellar population properties within galaxies and examine how galaxy morphology relates to gas content and star formation activity.

This paper is organized as follows. In Section \ref{sec:data}, we introduce the sample and optical datasets used. Section \ref{sec:analysis} outlines our analysis methodology, including image reduction, isophotal analysis, and structural decomposition. In Section \ref{sec:results}, we compare the derived quantities with star formation and \hi gas properties. Finally, we discuss the main results in Section \ref{sec:discuss} and present our summary in Section \ref{sec:summary}.

Throughout this paper, we adopt a standard $\Lambda$CDM cosmology with $\it {H}_{\rm 0} {\rm= 70\ km\ s^{-1}\ Mpc^{-1}}$, $\rm {\Omega_M = 0.3}$, and $\rm{\Omega_{\Lambda} = 0.7}$.

\section{Sample and Data}
\label{sec:data}

\subsection{xGASS}
In this work, we use galaxies from the extended GALEX Arecibo SDSS Survey \citep[xGASS;][]{Catinella2010, Catinella2018}, which includes 1179 galaxies with stellar masses $10^9 < M_* < 10^{11.5} M_\odot$ and redshifts $0.01 < z < 0.05$, representative of present-day galaxy properties. xGASS galaxies are uniformly selected from overlapping fields of the SDSS spectroscopic survey \citep{Abazajian2009}, the {\it Galaxy Evolution Explorer} (GALEX) imaging survey \citep{Martin2005}, and ALFALFA survey footprints. xGASS was designed to measure the atomic hydrogen (\hi) gas content using the Arecibo radio telescope, with each target observed until \hi emission was detected or the assigned gas mass fraction limit was reached. In the full sample, 375 galaxies have non-detections for \hi; thus, 5$\sigma$ upper limits are provided for their \hi masses, corresponding to gas fractions of a few percent.

In addition to \hi observations, a subset of approximately 500 xGASS galaxies was also observed with the IRAM 30m telescope to explore molecular gas content and provide H$_2$ estimates \citep{Saintonge2011, Saintonge2017}. Of these, 290 galaxies (about 60\%) were detected in molecular emission, and 3$\sigma$ upper limits for H$_2$ mass were provided for galaxies with no CO detections.

For each xGASS target, stellar mass and other optical parameters were derived from the SDSS DR7 MPA/JHU catalog \citep{Brinchmann2004}. Total SFR was measured using multiwavelength observations and corrections, combining GALEX near-UV and Wide-field Infrared Survey Explorer (WISE) mid-infrared (MIR) photometry \citep{Wright2010} or using the spectral energy distribution fitting technique \citep{Janowiecki2017}.

\begin{figure}
    \centering
    \includegraphics[width=\hsize]{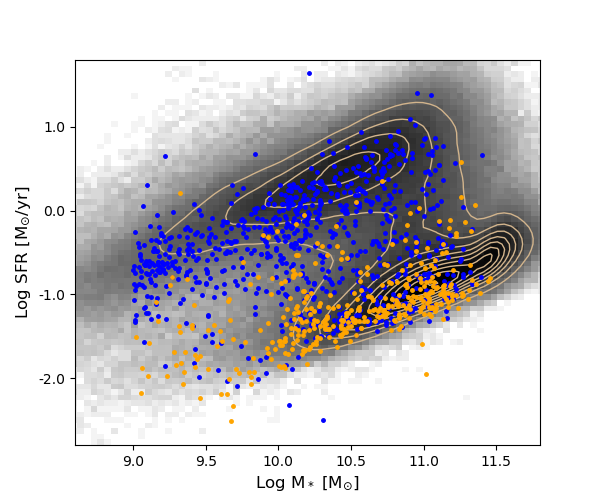}
	\caption{Distribution of galaxies in the xGASS sample in the SFR–$M_*$ plane. \him-detected galaxies are shown as blue points, while \hi non-detections are marked in orange. For comparison, the gray background represents the distribution of galaxies from the SDSS DR7 MPA/JHU catalog, with number density contours overlaid to highlight the overall population trends.
	\label{fig0}}
\end{figure}

Figure \ref{fig0} presents the distribution of our sample in the SFR-$M_*$ plane, , overlaid with galaxies from the SDSS DR7 MPA/JHU catalog for comparison. The figure demonstrates that our sample is uniformly distributed across the entire SFR-$M_*$ plane, ensuring a comprehensive representation of different galaxy populations.

\subsection{DESI Legacy Imaging Surveys}
We use optical photometry from Data Release 10 (DR10) of the DESI Legacy Imaging Survey \citep{Dey2019} to conduct structural decomposition and analysis for the xGASS sample. The Legacy Surveys, constructed from three ground-based telescopes, combine data from three wide-area optical imaging surveys: the Dark Energy Camera Legacy Survey (DECaLS), the Mayall $\it z$-band Legacy Survey (MzLS), and the Beijing–Arizona Sky Survey \citep[BASS;][]{Zou2017}. Together, these surveys cover approximately 14,000 $\rm deg^2$ of the northern extragalactic sky in the $g$, $r$, and $z$ optical bands, with nearly uniform depths reaching 5$\sigma$ limits of $g$ = 24.0, $r$ = 23.4, and $z$ = 22.5 AB mag for faint galaxies. The typical point-spread function (PSF) is around $1^{''}$, though it varies across bands and surveys. Additionally, DESI Legacy Imaging Survey data releases include MIR photometry from WISE and NEOWISE \citep{Mainzer2014}. Using Tractor, an inference-based forward-modeling approach, the Legacy Surveys determine source positions, shape parameters, and photometry to produce the final source catalogs.

\begin{deluxetable*}{ccccccccccccccccc}
\setlength{\tabcolsep}{3pt}
	\tablecaption{Bulge and Disk Properties from Structure Decomposition of Galaxies \label{tab1}}
	\centering
	\small
	\tablehead{
		\colhead{GASS} & \colhead{RA} & \colhead{DEC} & \colhead{$g$} & \colhead{$r$} & \colhead{$z$} & \colhead{$ n_{global}$} & \colhead{$R_e$} & \colhead{$g_d$} & \colhead{$r_d$} & \colhead{$z_d$} & \colhead{$R_{e,d}$} & \colhead{$g_b$} & \colhead{$r_b$} & \colhead{$z_b$} & \colhead{$R_{e,b}$} & \colhead{$n_{b}$}\\
		\colhead{} & (J2000.0) & (J2000.0) & \colhead{(mag)} & \colhead{(mag)} & \colhead{(mag)} & \colhead{} & \colhead{(arcsec)} & \colhead{(mag)} & \colhead{(mag)} & \colhead{(mag)} & \colhead{(arcsec)} & \colhead{(mag)} & \colhead{(mag)} & \colhead{(mag)} & \colhead{(arcsec)} & \colhead{}}
\decimalcolnumbers
\startdata
3151 &  7.92941 & 14.48647 & 16.4 & 15.5 & 14.9 & 4.7 & 4.9 & 15.9 & 17.0 & 15.4 & 21.8 & 16.7 & 16.8 & 16.8 & 20.8 & 6.2 \\
3157 &  7.63725 & 14.94319 & 16.2 & 15.3 & 14.7 & 4.1 & 4.1 & 17.0 & 18.7 & 16.0 & 18.4 & 15.7 & 16.3 & 15.0 & 17.4 & 4.6 \\
3163 &  7.19513 & 15.09150 & 16.0 & 15.2 & 14.5 & 0.8 & 5.6 & 15.3 & 16.1 & 14.6 & 18.7 & 23.4 & 25.0 & 23.6 &  3.3 & 2.1 \\
3189 & 10.09787 & 14.61375 & 16.7 & 16.0 & 15.4 & 0.5 & 8.5 & 16.9 & 17.4 & 16.2 & 25.1 & 25.4 & 26.4 & 25.6 &  2.4 & 3.9 \\
3233 & 12.89187 & 13.91464 & 16.5 & 15.8 & 15.2 & 1.2 & 2.3 & 16.4 & 17.3 & 15.9 & 12.7 & 16.5 & 17.2 & 15.9 &  7.4 & 0.9 \\
3258 & 13.32064 & 16.09892 & 17.3 & 16.3 & 15.5 & 1.4 & 4.2 & 16.4 & 17.5 & 15.7 & 13.3 & 22.7 & 23.4 & 22.4 &  1.6 & 0.5 \\
3261 & 13.88588 & 15.77582 & 16.3 & 15.6 & 15.1 & 1.4 & 3.1 & 16.2 & 16.7 & 15.7 & 16.9 & 17.9 & 18.5 & 17.5 &  9.2 & 2.0 \\
3284 & 15.72436 & 14.19455 & 16.4 & 15.5 & 14.7 & 1.6 & 8.0 & 15.8 & 16.7 & 15.1 & 25.1 & 18.2 & 19.7 & 17.1 &  7.2 & 0.9 \\
3301 & 14.81382 & 13.57594 & 15.5 & 14.6 & 13.9 & 4.0 & 6.4 & 15.1 & 16.2 & 14.3 & 25.0 & 16.2 & 16.4 & 15.9 & 24.0 & 7.4 \\
3321 & 15.61839 & 15.74917 & 16.8 & 15.6 & 14.6 & 1.0 & 9.0 & 15.9 & 17.0 & 14.9 & 24.8 & 18.1 & 22.5 & 17.4 & 23.8 & 0.5 \\
\enddata
	\tablecomments{Column (1): Galaxy IDs in xGASS. Column (2)-(3): J2000 coordinates in degree. Colum (4)-(8): $grz$ magnitudes, S\'{e}rsic index and effective radius for global galaxy from DESI LS. Colum (9)-(12): $grz$ magnitudes and effective radius for disks from structure decomposition. Colum (13)-(17): $grz$ magnitudes, effective radius and S\'{e}rsic index for bulges from structure decomposition. The full version of this table is available at ScienceDB: \dataset[10.57760/sciencedb.25139]{\doi{10.57760/sciencedb.25139}}.
	}
\end{deluxetable*}


\section{Analysis}
\label{sec:analysis}
In this study, we use $g$-, $r$-, and $z$-band images from the DESI Legacy Imaging Surveys to conduct structural analysis on xGASS galaxies. For each galaxy, coadded images for each band are obtained through the cutout service on the Legacy Surveys website, along with the average, inverse-variance-weighted pixelized PSF for each bandpass at the field’s center. To ensure the full field of view for each target, the image cutout is set to a size of 8 times the galaxy’s half-light radius (${\it shape\_r}$) and centered on the galaxy's position. A fixed pixel scale of 0.262 arcsec/pixel is used across all three bands.

\subsection{Image Masking and Cleaning}
To achieve accurate measurements, it is essential to identify and mask signals from unrelated objects, such as bright foreground stars and background galaxies near each target galaxy. For this purpose, we adopt the object masking method described by \citet{Zhou2020}. The specific steps are as follows.

First, we create a ``flag image'' to mark the pixels affected by unrelated objects. Using SExtractor \citep{Bertin1996}, we detect objects in each image and generate both a photometric catalog and a segmentation image for each field. To avoid misidentifying pixels from the main body of the target galaxy, we carefully test SExtractor's input parameters and select pixels for masking based on their proximity to the target galaxy. These masked pixels are then used to construct the flag image.

After defining the initial masking regions, we expand these regions outward by several pixels to ensure that we cover the faint outer halos of bright contaminating objects. Once the flag image is generated, we replace the masked regions by linearly interpolating pixel intensity values from nearby unaffected regions to minimize interference from unrelated sources. We use the r-band image as the reference to create the flag image, applying this same flag image to all three bands ($g$, $r$, $z$) owing to their similar data quality.

\subsection{Isophotal analysis}
Using the star-cleaned images, we apply the ASTROPY package PHOTUTILS to perform isophotal analysis. Elliptical isophotes are fitted and measured with the isophote.Ellipse function in PHOTUTILS to derive 1D surface brightness profiles and determine the radial variation of geometric parameters, such as ellipticity, position angle, and color gradients. These geometric parameters also allow us to identify large-scale stellar bars in galaxies.

For each galaxy, we first perform ellipse fitting on the $r$-band image, allowing all parameters to vary freely. We use the initial values for center coordinates, ellipticity, and position angle from the photometric catalogs of the DESI Legacy Imaging Surveys. The fitting step is set to 1 pixel, and the maximum semimajor axis length is limited to three times the half-light radius. We then apply the r-band fitting solution to the g-band and z-band images, using the same position and shape parameters to construct color profiles.

Stellar bars can be identified through the isophotal profiles of ellipticity and position angle \citep{Jogee2004, Erwin2005, Gadotti2008}. In barred galaxies, ellipticity generally increases steadily within the bar region, reaching a maximum above 0.25 and then dropping by more than 0.1 at the bar’s end. The position angle remains nearly constant within the bar region and then shifts abruptly by more than $10^{\circ}$ within 1$\arcsec$-2$\arcsec$ beyond the bar. We further validate the identified barred galaxies through visual inspection, correcting obvious misidentifications and recovering bars that may have been missed. This process results in 312 detected barred galaxies, approximately one-third of the sample. It is important to note that this method primarily selects strong bars.

\subsection{Structure decomposition}
To further characterize the structural properties of the sample, we perform 2D multi-component decompositions using GALFIT \citep{Peng2010} and GALFITM \citep{Vika2013, Haussler2022}, estimating the light ratio between the bulge and disk components.

The fitting process for each galaxy involves four iterative steps. First, we use GALFIT to fit a single S\'{e}rsic model with a free S\'{e}rsic index {\it n} based on the r-band image of each target. Initial parameters—such as magnitude, S\'{e}rsic index, effective radius (in pixels), axis ratio, and position angle—are taken from the photometric results of the DESI Legacy Imaging Surveys. The initial center position is set to the center of each image cutout. Since the sky background was removed during data reduction, resulting in a near-zero constant sky level, no additional background component is included in the fitting.

Next, we perform a two-component fit, consisting of an exponential disk (fixed S\'{e}rsic index {\it n} = 1) and a bulge model (with {\it n} free). The parameter values derived from the initial single-component fit are used as starting values for each component in this step. For barred galaxies, we introduce an additional free S\'{e}rsic model in the third step to account for the stellar bar. Finally, we use GALFITM to simultaneously fit the {\it grz} bands, using the parameters from the previous step as initial values. In this stage, we set the r-band image as the reference, constraining the shape parameters of each model to remain consistent across all three bands while allowing the magnitude to vary freely.

\begin{figure*}
    \centering
    \includegraphics[width=0.55\hsize]{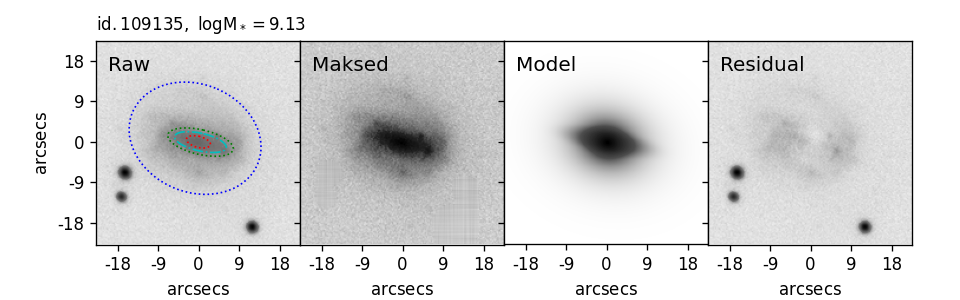}
    \includegraphics[width=0.44\hsize]{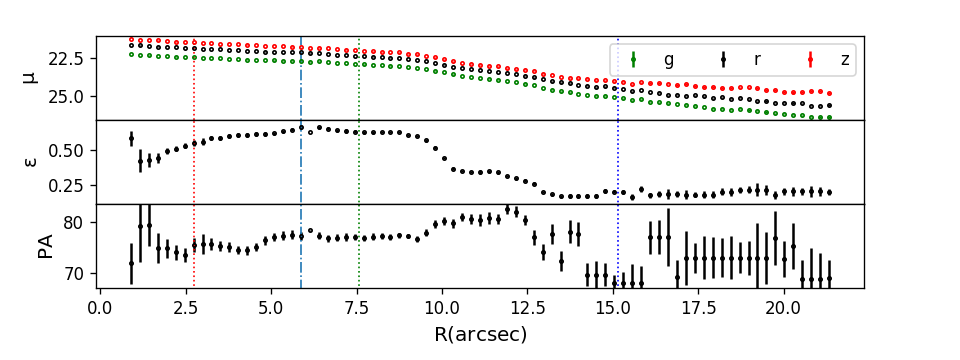}
    \includegraphics[width=0.55\hsize]{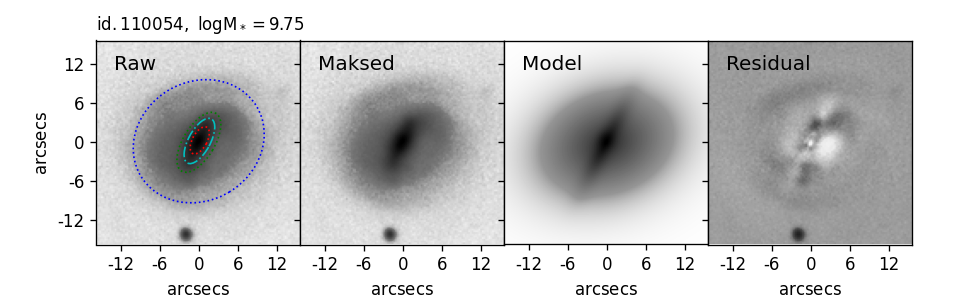}
    \includegraphics[width=0.44\hsize]{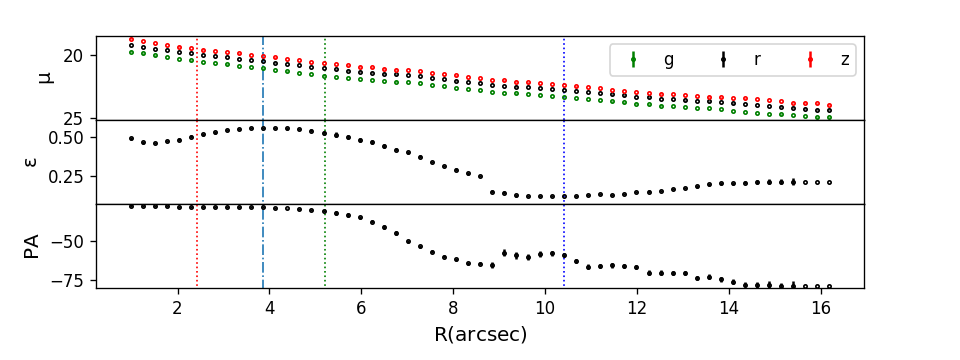}
    \includegraphics[width=0.55\hsize]{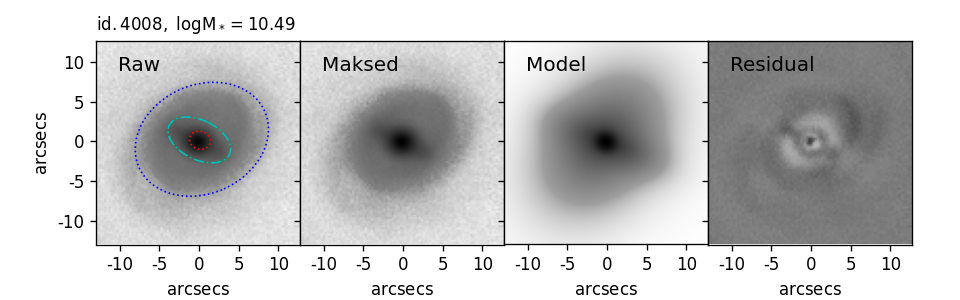}
    \includegraphics[width=0.44\hsize]{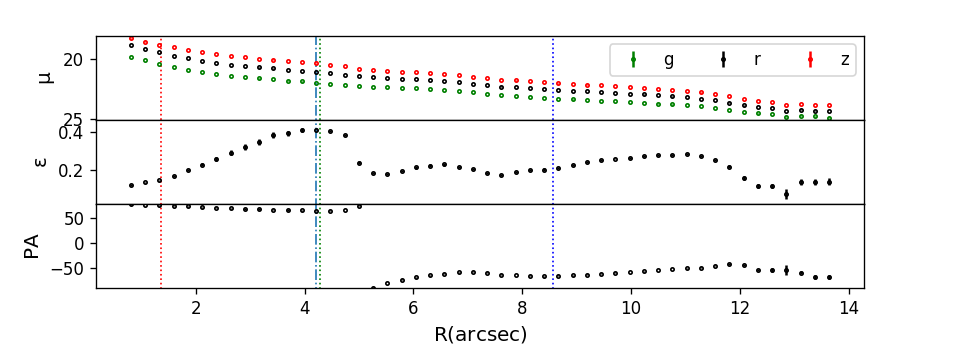}
    \includegraphics[width=0.55\hsize]{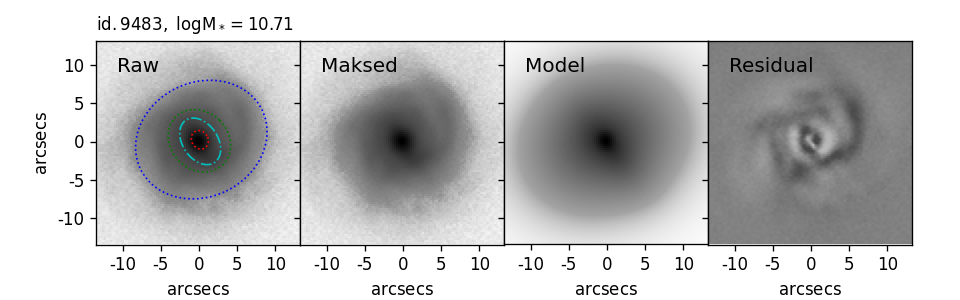}
    \includegraphics[width=0.44\hsize]{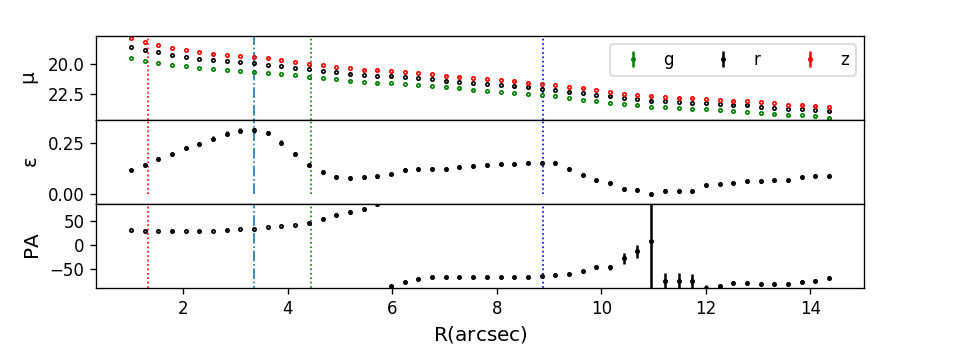}
    \includegraphics[width=0.55\hsize]{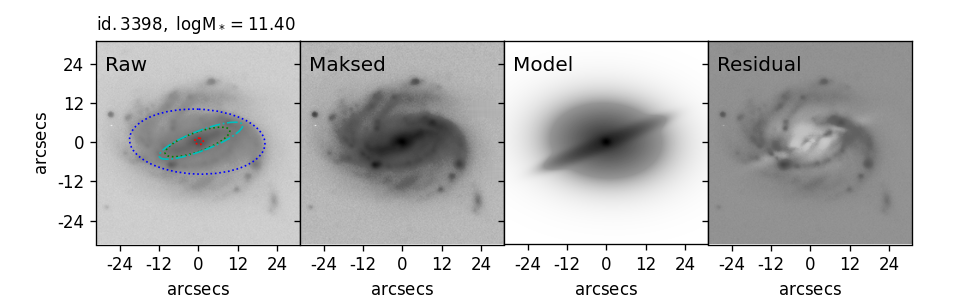}
    \includegraphics[width=0.44\hsize]{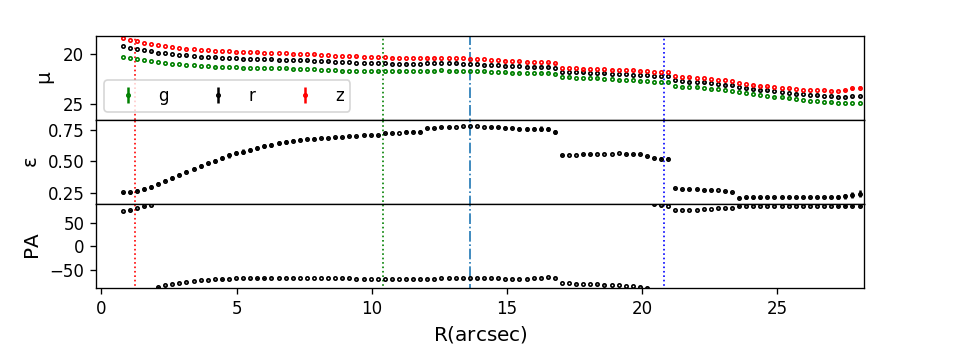}
	\caption{Example illustration of 2D image decomposition and isophotal analysis for barred galaxies across different stellar masses. The galaxy ID and stellar mass are indicated at the top of the left panel. Each row presents the fitting results for a specific galaxy. The {\it left} panel displays the raw $\it r$-band cutout, star-cleaned image, and the 2D model and residual images from GALFIT fitting, which includes bulge, disk, and bar components. The {\it right} panel shows the radial profiles of surface brightness in $\it grz$ bands, along with ellipticity ($\it e$) and position angle (PA) from isophotal analysis based on the $\it r$-band (top to bottom). In both panels, dotted lines indicate three key regions within the galaxy: central 1 kpc (red), the effective radius $R_{\rm eff}$ (green), and 2$R_{\rm eff}$ (blue). Dashed-dotted lines mark the extent of the bar, indicated by a peak in ellipticity and a stable PA.
	\label{fig1}}
\end{figure*}

\begin{figure*}
    \centering
    \includegraphics[width=0.55\hsize]{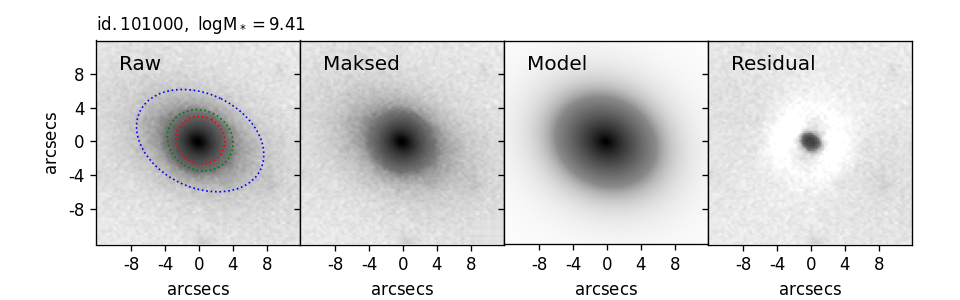}
    \includegraphics[width=0.44\hsize]{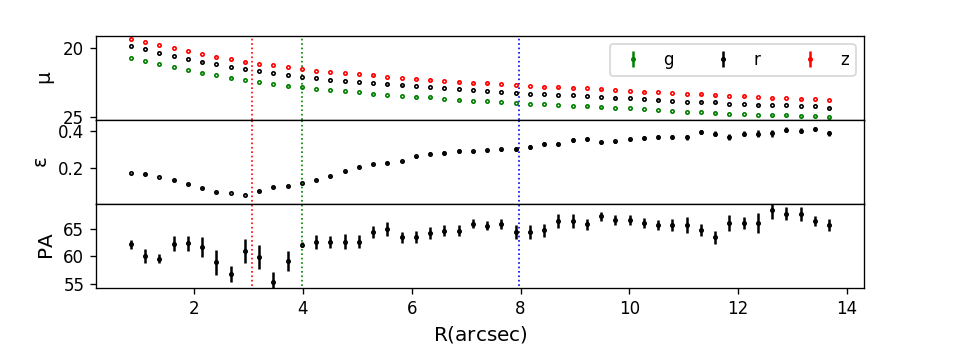}
    \includegraphics[width=0.55\hsize]{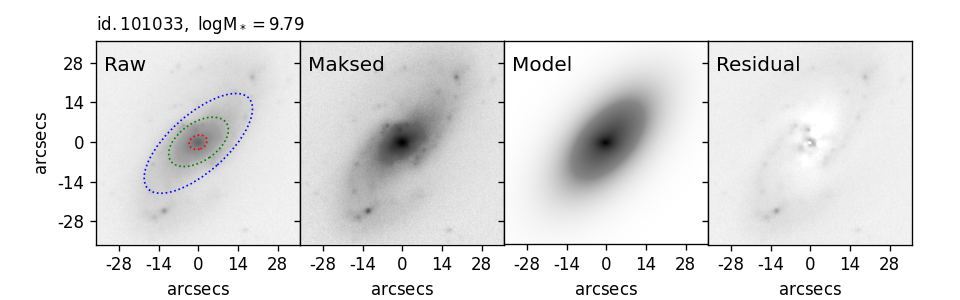}
    \includegraphics[width=0.44\hsize]{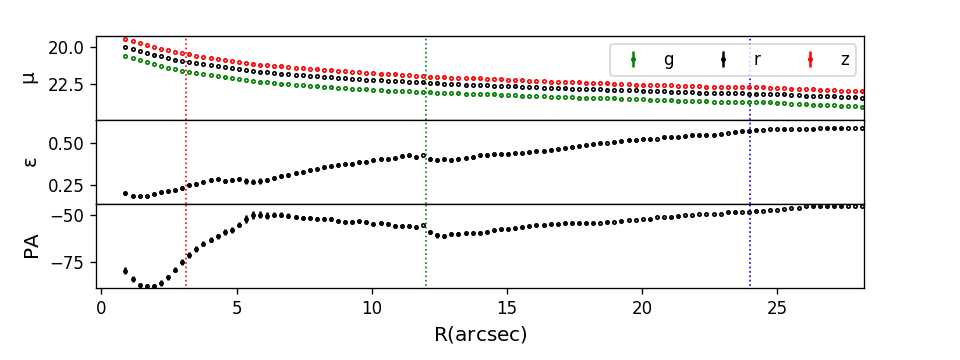}
    \includegraphics[width=0.55\hsize]{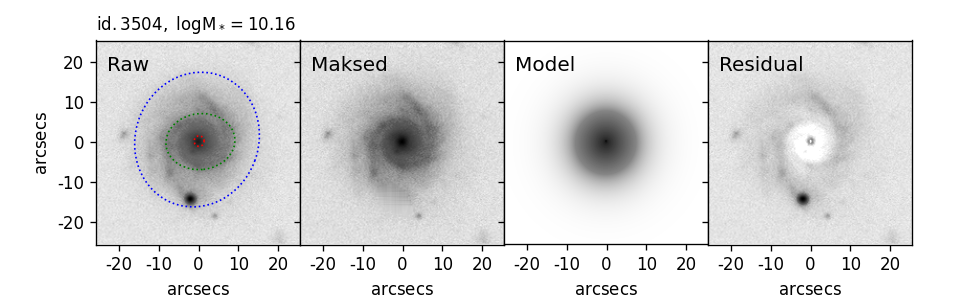}
    \includegraphics[width=0.44\hsize]{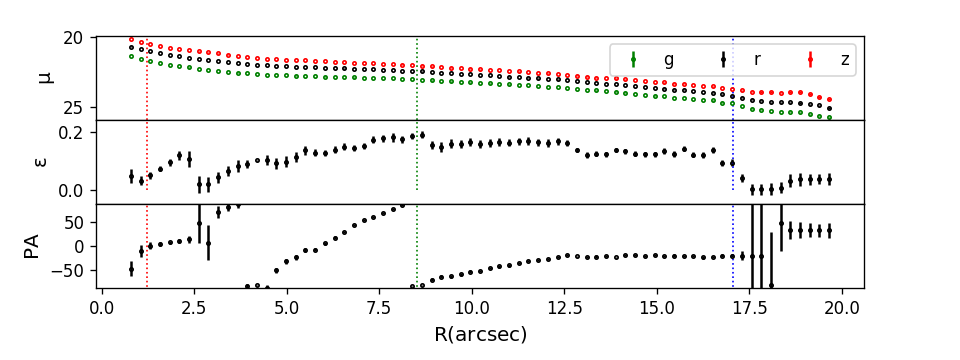}
    \includegraphics[width=0.55\hsize]{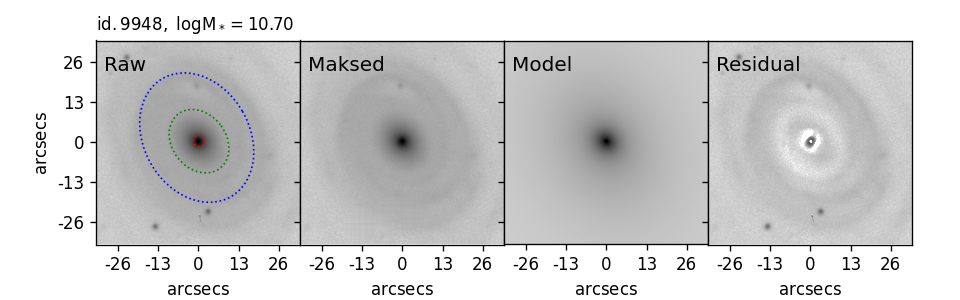}
    \includegraphics[width=0.44\hsize]{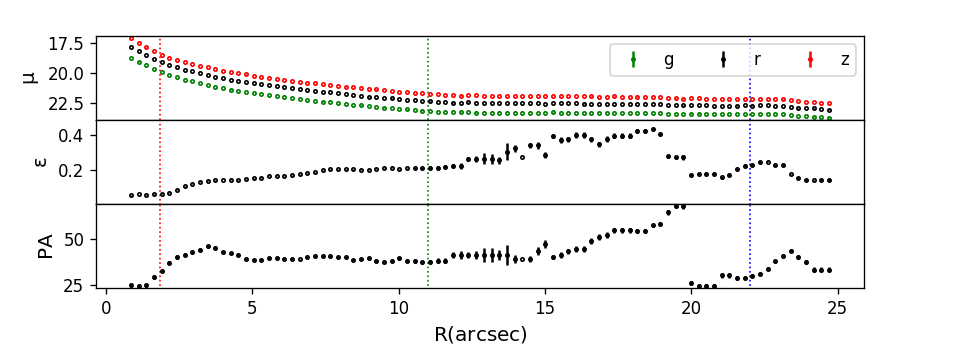}
    \includegraphics[width=0.55\hsize]{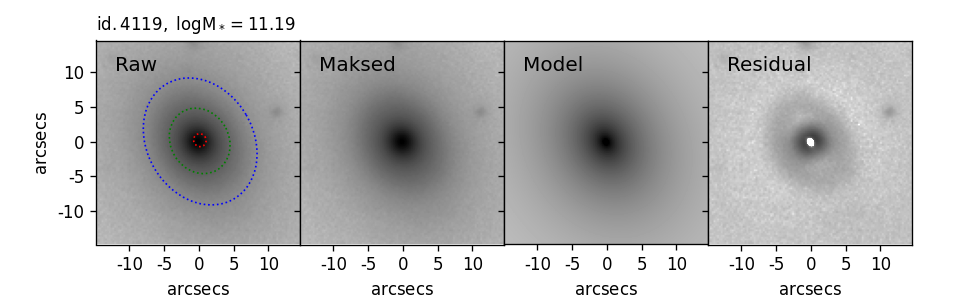}
    \includegraphics[width=0.44\hsize]{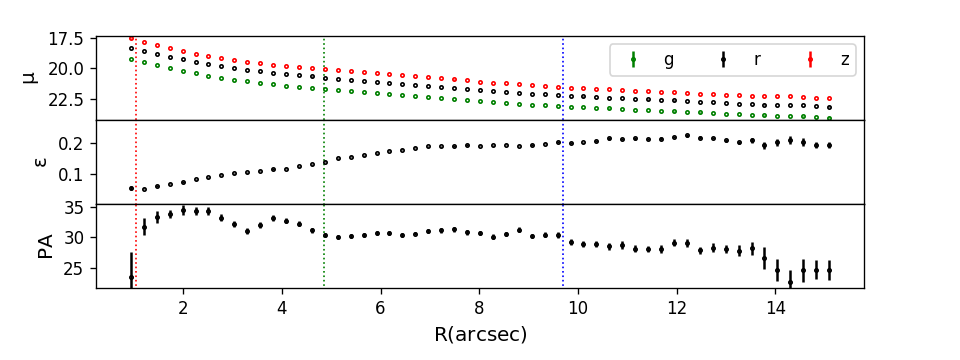}
	\caption{Similar to Figure~\ref{fig1}, but illustrating non-barred galaxies across different stellar masses. Each galaxy is decomposed into bulge and disk components.
	\label{fig2}}
\end{figure*}

\begin{longrotatetable}
\begin{deluxetable}{ccccccccccccccc}
\tabletypesize{\footnotesize}
\tablecaption{Color Properties of Galaxies from Isophotal Analysis \label{tab3}}
\centering
\small
\tablehead{
		\colhead{GASS} & \colhead{$(g-r)_{1kpc}$} & \colhead{$(r-z)_{1kpc}$} &  \colhead{$(g-r)_{1re}$} & \colhead{$(r-z)_{1re}$} & \colhead{$(g-r)_{2re}$} & \colhead{$(r-z)_{2re}$} 
		& \colhead{$(g-r)_{1re^{'}}$} & \colhead{$(r-z)_{1re^{'}}$} & \colhead{$(g-r)_{2re^{'}}$} & \colhead{$(r-z)_{2re^{'}}$}
		& \colhead{$\nabla_{g-r,1re}$} & \colhead{$\nabla_{r-z,1re}$} & \colhead{$\nabla_{g-r,2re}$} & \colhead{$\nabla_{r-z,2re}$}}
\decimalcolnumbers
\startdata
3151 & 1.06 & 0.63 & 0.87 & 0.59 & 0.80 & 0.56 & 0.72 & 0.55 & 0.70 & 0.53 & -0.15 & -0.05 & -0.02 & -0.02 \\
3157 & 0.85 & 0.68 & 0.85 & 0.65 & 0.82 & 0.62 & 0.82 & 0.61 & 0.78 & 0.59 & -0.08 & -0.06 & -0.05 & -0.02 \\
3163 & 0.94 & 0.69 & 0.84 & 0.63 & 0.78 & 0.59 & 0.78 & 0.59 & 0.71 & 0.53 & -0.09 & -0.08 & -0.07 & -0.06 \\
3189 & 0.80 & 0.51 & 0.65 & 0.44 & 0.58 & 0.40 & 0.55 & 0.38 & 0.53 & 0.39 & -0.21 & -0.18 & -0.02 &  0.01 \\
3233 & 0.70 & 0.54 & 0.67 & 0.50 & 0.68 & 0.48 & 0.64 & 0.45 & 0.74 & 0.49 & -0.04 & -0.08 &  0.10 &  0.05 \\
3258 & 1.01 & 0.72 & 0.95 & 0.68 & 0.89 & 0.66 & 0.91 & 0.66 & 0.80 & 0.62 & -0.08 & -0.03 & -0.11 & -0.05 \\
3261 & 0.64 & 0.39 & 0.61 & 0.38 & 0.57 & 0.37 & 0.57 & 0.38 & 0.52 & 0.37 & -0.11 &  0.01 & -0.05 & -0.01 \\
3284 & 1.06 & 0.90 & 0.93 & 0.75 & 0.87 & 0.68 & 0.85 & 0.64 & 0.79 & 0.59 & -0.16 & -0.19 & -0.06 & -0.05 \\
3301 & 1.04 & 0.69 & 0.91 & 0.65 & 0.87 & 0.65 & 0.81 & 0.63 & 0.82 & 0.66 & -0.12 &  0.02 &  0.01 &  0.03 \\
3321 & 1.04 & 0.87 & 1.05 & 0.87 & 1.01 & 0.83 & 1.03 & 0.84 & 0.94 & 0.75 & -0.07 & -0.09 & -0.09 & -0.09 \\
\hline
\enddata
\tablecomments{Column (1): Galaxy IDs in xGASS. Colum (2)-(7): Gloabl colors ($g-r$) and ($r-z$) within the radius of 1 Kpc, $R_e$ and 2$R_e$ of galaxies.  Colum (8)-(11): Colors ($g-r$) and ($r-z$) in the region of $1 kpc < R < 1R_e$ and $1 R_e < R < 2 R_e$. Colum (12)-(15): The ($g-r$) and ($r-z$) color gradients in inner and outer regions of galaxies. The full version of this table is available at ScienceDB: \dataset[10.57760/sciencedb.25139]{\doi{10.57760/sciencedb.25139}}}
\end{deluxetable}
\end{longrotatetable}


Figures \ref{fig1} and \ref{fig2} show examples of the 1D profiles and 2D decomposition fitting results for barred and non-barred galaxies across different stellar masses, respectively. Based on the 1D geometric profile, when a typical stellar bar is detected in a galaxy, the decomposition analysis fits three components: bulge, disk, and bar. Otherwise, a bulge+disk decomposition is fitted. To illustrate key regions of galaxies, three reference positions are marked with dotted lines in both figures: central 1 kpc, the effective radius $R_{\rm e}$, and $2R_{\rm e}$. For barred galaxies, the bar region is highlighted with dashed-dotted lines, identified based on a peak in ellipticity and a stable position angle.

Tables \ref{tab1}, \ref{tab2}, and \ref{tab3} summarize the main properties of the xGASS galaxies used or presented in this study. Table \ref{tab1} includes the results of bulge and disk structure decomposition, along with photometric data sets retrieved from the DESI Legacy Imaging Survey. Table \ref{tab2} presents the bar properties from the structure decomposition of 312 barred galaxies in the sample. Table \ref{tab3} shows the color properties of the sample, derived from isophotal analysis.

\begin{deluxetable}{cccccc}
\setlength{\tabcolsep}{3pt}
	\tablecaption{Bar Properties from Structure Decomposition of Galaxies \label{tab2}}
	\centering
	\tablehead{
		\colhead{GASS} & \colhead{$g_{bar}$} & \colhead{$r_{bar}$} & \colhead{$z_{bar}$} & \colhead{$R_{e,bar}$} & \colhead{$b/a_{bar}$}\\
		\colhead{} & \colhead{(mag)} & \colhead{(mag)} & \colhead{(mag)} & \colhead{(arcsec)} & \colhead{}}
\decimalcolnumbers
\startdata
3189 & 17.4 & 18.7 & 17.1 & 23.1 & 0.4 \\
3261 & 16.9 & 17.4 & 16.4 & 11.6 & 0.5 \\
3398 & 16.0 & 17.1 & 15.4 & 40.1 & 0.2 \\
3439 & 17.6 & 18.8 & 17.1 &  8.8 & 0.5 \\
3519 & 16.9 & 17.6 & 16.5 & 35.6 & 0.4 \\
3634 & 16.2 & 17.4 & 15.6 & 16.9 & 0.6 \\
3645 & 17.0 & 18.0 & 16.4 &  9.4 & 0.4 \\
3777 & 17.2 & 18.1 & 16.7 & 11.2 & 0.6 \\
3817 & 18.9 & 19.7 & 18.6 & 18.6 & 0.4 \\
3957 & 18.6 & 19.6 & 18.1 &  7.8 & 0.5 \\
\enddata
	\tablecomments{Column (1): Galaxy IDs in xGASS. Colum (2)-(4): $grz$ magnitudes for bars. Colum (5): effective radius of bars. Colum (6): axis ratio of bars. The full version of this table is available at ScienceDB: \dataset[10.57760/sciencedb.25139]{\doi{10.57760/sciencedb.25139}}}
\end{deluxetable}


\section{Results}
\label{sec:results}
In this section, we investigate the correlations between bulge/disk structures and the star formation properties of the sample galaxies. We also examine the dependence of \hi gas content on galactic structures. The primary goal of this analysis is to enhance our understanding of how structural components influence galaxy evolution. To quantify these correlations, we also calculate the Pearson correlation coefficient with associated $p$-value for all the parameter pairs investigated in the following figures. These valuse are summarized in Table \ref{t_pearson}.

\subsection{Structural Properties versus star formation}
As shown in previous observational studies, the Main Sequence (MS) relation reveals a strong link between the SFR and stellar mass ($M_*$) in star-forming galaxies \citep{Speagle2014, Chang2015, Tacchella2016}. To explore secondary dependencies in star formation, we examine correlations between star formation and various structural components. We define the distance from the star-forming MS (SFMS) as $\rm \Delta SFR_{MS} = \log(SFR) - \log(SFR_{MS})$, where $\rm \log(SFR_{MS})$ is calculated using the following definition from \citet{Renzini2015}:
\begin{eqnarray}
	{\rm log\ SFR_{MS}}(M_{\odot}~{\rm yr}^{-1}) = 0.76 \times {\rm Log}\ {M(M_{\odot})} - 7.64.
\end{eqnarray}

\subsubsection{Structural fraction v.s. star formation}
\begin{figure*}
    \centering
    \includegraphics[width=\hsize]{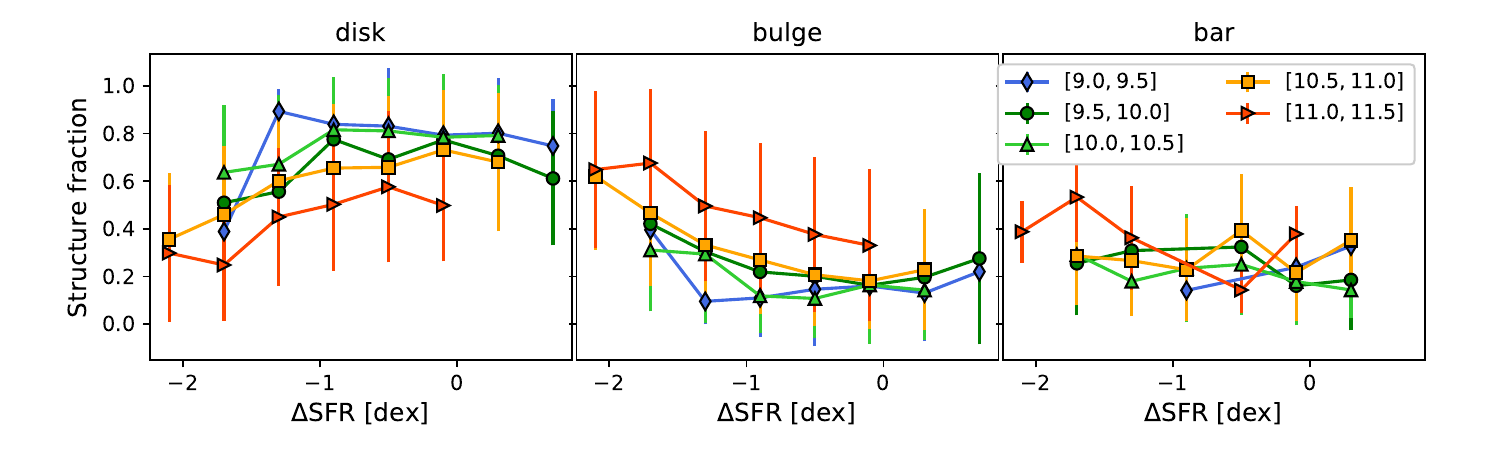}
	\caption{Relationship between structural flux fractions and the offset from the star formation main sequence (\dSFRMS). From left to right, panels show the disk, bulge, and bar fractions, respectively. In each panel, the sample is divided into five stellar mass subclasses in logarithmic bins of 0.5 dex, ranging from $10^9$ to $10^{11.5}$ $M_{\odot}$, as indicated in the legend on the right. Only bins including at least 5 galaxies are shown. Points and error bars represent the median values and corresponding 1$\sigma$ scatter in each bin.
	\label{fig3}}
\end{figure*}

Figure \ref{fig3} illustrates how the bulge, disk, and bar fractions vary with $\rm \Delta SFR_{MS}$ for the sample galaxies. We divide the galaxies into five stellar mass subclasses, each with a 0.5 dex width, ranging from $10^9$ to $10^{11.5} M_{\odot}$. Within each mass interval, the mean structural fractions are calculated for each star formation bin. Generally, more massive galaxies show a higher bulge fraction and lower disk fraction, particularly for galaxies with stellar masses above $10^{11} M_{\odot}$. The bulge fraction typically decreases with increasing $\rm \Delta SFR_{MS}$ within each stellar mass bin, indicating that star-forming galaxies tend to be disk-dominated, while quiescent systems are predominantly bulge-dominated. 
For the bar fraction in the right panel of Figure \ref{fig3}, there is not significant variation across different stellar mass bins and $\rm \Delta SFR_{MS}$ values within the associated uncertainties, along with the Pearson correlation coefficient of $c \sim$ 0.04 and $p$-value $>$ 0.1.

In Figure \ref{fig3}, we further observe a mild upturn in the median bulge fraction with $\rm \Delta SFR_{MS} > 0$, particularly in galaxies with stellar masses below $10^{11} M_{\odot}$. For instance, the median bulge fraction decreases from ~0.4 at $\rm \Delta SFR_{MS}$ = -2.0 to -1.5, to ~0.3 at $\rm \Delta SFR_{MS}$ = -1.5 to -1.0. It reaches the lowest value of ~0.16 near $\rm \Delta SFR_{MS}$ $\approx$ 0 and then increases to ~0.3 for $\rm \Delta SFR_{MS} >$ 0.5. 
This trend is consistent with previous findings, where disk-dominated galaxies typically lie along the SFMS, bulge-dominated systems are found in the quiescent region, and galaxies in the Green Valley or above the SFMS often exhibit intermediate bulge fractions \citep{Wuyts2011, Morselli2017}.

Nevertheless, compared to \citet{Morselli2017}, our sample includes fewer galaxies, particularly at the high-mass, high-SFR end. As shown in Figure \ref{fig3}, the observed trend is accompanied by considerable scatter, with typical 1$\sigma$ uncertainties of 0.15–0.20 per bin. Across stellar mass bins at fixed $\rm \Delta SFR_{MS}$, median differences in bulge fraction fall within this scatter, indicating that the trend is relatively weak.

Moreover, \citet{Cook2020}, who analyzed the same xGASS sample, also did not find a similar upturn trend in bulge fraction, but a monotonic decrease from the passive population to the upper envelope of the SFMS.
Unlike the use of GALFIT in this study, \citet{Cook2020} employed $\it PROFIT$ \citep{Robotham2017}, a Bayesian fitting tool for modeling 2D photometric galaxy profiles. For each galaxy, they applied complex bulge $+$ disk models and assumed a S\'{e}rsic bulge and a near-exponential disk but did not account for bar structures. 
In addition to methodological differences, \citet{Cook2020} examined trends at both fixed stellar mass and fixed SFR. These differences in methodological and analysis methods may contribute to variations in results. Additionally, the relatively small sample size likely amplifies these differences. Therefore, a more comprehensive analysis using larger galaxy samples and different algorithms will be necessary in future work.

\begin{figure}
    \centering
    \includegraphics[width=\hsize]{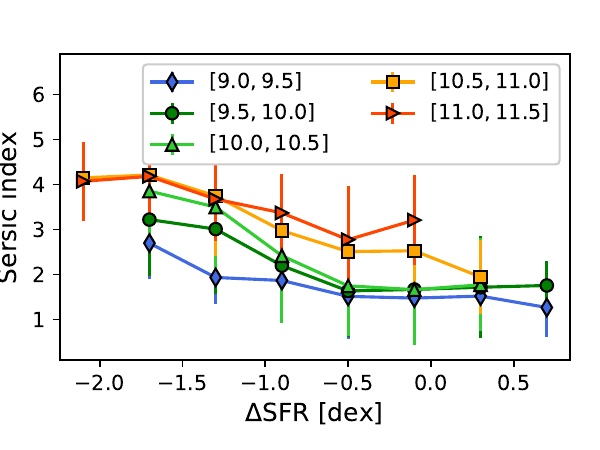}
	\caption{Relationship between global S\'{e}rsic index $\it n$ and \dSFRMS. Similar to Figure \ref{fig3}, the sample is divided into five stellar mass subclasses with logarithmic bins, as indicated in the legend.
	\label{fig4}}
\end{figure}

In Figure \ref{fig4}, we further use the global S\'{e}rsic index {\it n} as an additional indicator of surface brightness profiles and structural characteristics of galaxies. We analyze the distribution of {\it n} as a function of $\rm \Delta SFR_{MS}$ across different stellar mass bins. Generally, galaxies with higher S\'{e}rsic indices exhibit larger bulge-to-total ratios and higher central stellar mass densities, suggesting they are bulge-dominated \citep{Vulcani2014, Shibuya2015}. Consistent with Figure \ref{fig3} and prior studies \citep[e.g.,][]{Wuyts2011, Stephenson2024}, we find that galaxies in higher stellar mass bins tend to have larger S\'{e}rsic indices on average, while those in lower mass bins generally have smaller S\'{e}rsic indices. In each mass bin, more quiescent galaxies (i.e., those with smaller $\rm \Delta SFR_{MS}$) show higher S\'{e}rsic indices. This trend is more pronounced for massive, quiescent galaxies, whereas for star-forming galaxies with stellar masses below $\sim 10^{10} M_{\odot}$, the S\'{e}rsic index remains relatively constant within statistical errors.

\citet{Wuyts2011} also found a reversal of {\it n} distribution for starburst galaxies above MS, implying that high-SFR galaxies have enhanced surface brightness profiles or bulge-dominated morphologies. Nevertheless, this trend is not obvious in our result, likely due to the lack of rare starburst galaxies that form the high-SFR tail in the xGASS sample. 
The local sample in \citet{Wuyts2011} was drawn from the spectroscopic catalog of SDSS DR7. It spans a stellar mass range of $10^{8}-10^{12}\ M_{\odot}$ and an SFR range of $10^{-2}-10^{2}\ M_{\odot}\ yr^{-1}$. This broad coverage includes a wide variety of galaxies, from quiescent systems to extreme starbursts. In contrast, as shown in Figure \ref{fig0}, the xGASS sample contains fewer starburst galaxies and lacks galaxies at the highest and lowest stellar mass extremes. This is especially true for galaxies with $\rm \Delta SFR_{MS} > 0.5$. These differences in sample selection likely contribute to the variation in observed trends between the two studies.

\begin{figure*}
    \centering
    \includegraphics[width=0.9\hsize]{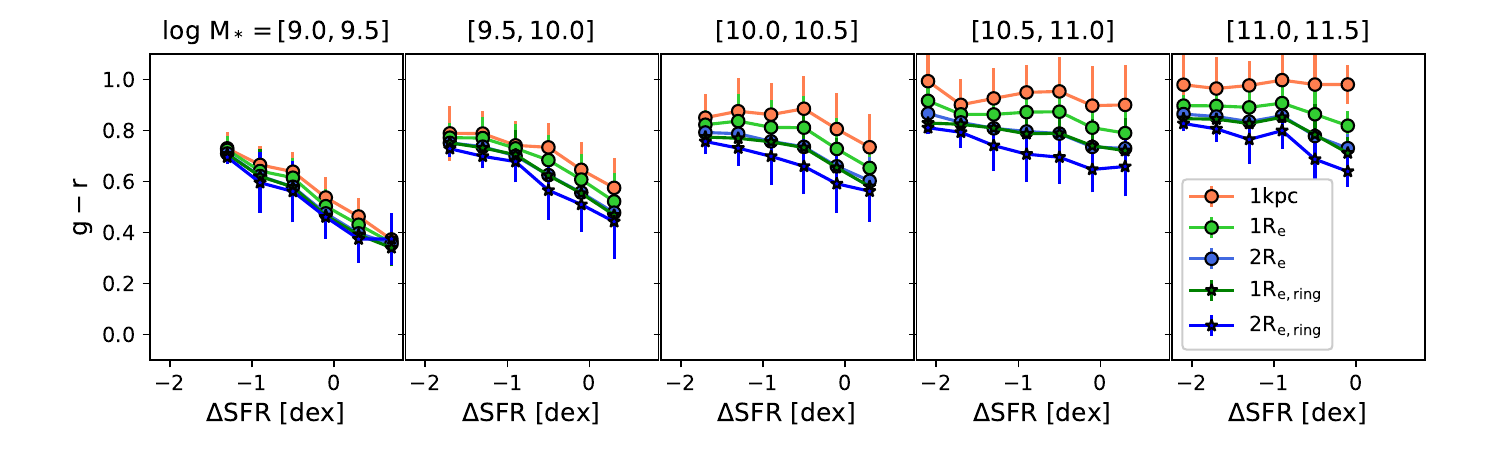}
	\caption{Relationship between galaxy color $g - r$ and \dSFRMS. From left to right, the panels correspond to five stellar mass bins, each with a width of 0.5 dex over the range $\rm M_* = 10^9 - 10^{11.5}\ M_{\odot}$, as in Figure \ref{fig3}. In each panel, symbols represent colors measured from different galactic regions: 1 kpc (orange circle) - $\rm R < 1\ kpc$, 1 $R_e$ (green circle) - $\rm R < 1R_e$, 2 $R_e$ (blue circle) - $\rm R < 2R_e$, 1 $R_e$ ring (green star) - $\rm 1\ kpc < R < 1R_e$, and 2 $R_e$ ring (blue star) - $\rm 1 R_e < R < 2 R_e$. Symbol definitions are in the legend of the right panel.
	\label{fig5}}
	\end{figure*}

\begin{figure*}
    \centering
    \includegraphics[width=0.9\hsize]{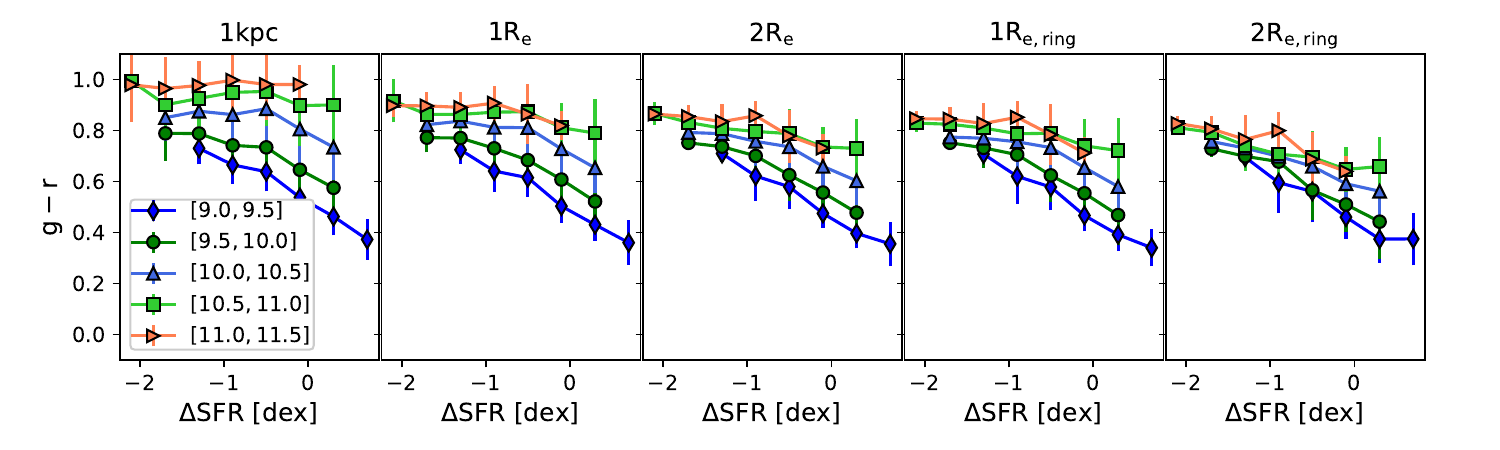}
	\caption{Relationship between galaxy color $g - r$ and \dSFRMS. Similar to Figure \ref{fig5}, but here, colors from the same galactic region across different stellar mass bins are grouped in each panel. Symbols representing different stellar mass bins are shown in the legend of the left panel.
		\label{fig6}}
\end{figure*}

\begin{figure*}
    \centering
    \includegraphics[width=0.9\hsize]{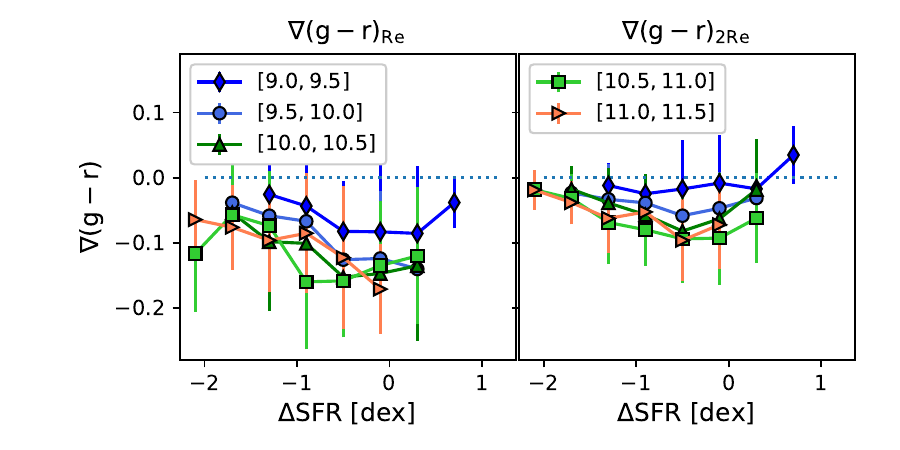}
	\caption{Relationship between radial color gradient and \dSFRMS. The left panel shows the color gradient index in the inner region of the galaxy ($\rm \nabla(g - r){re}$), while the right panel displays the color gradient index in the outer region ($\rm \nabla(g - r){2re}$). Each panel divides the sample into the same stellar mass bins as shown in Figure \ref{fig3}. The horizontal dotted lines are at $\rm \nabla(g - r) = 0$, separating the increasing and decreasing color gradients.
		\label{fig7}}
\end{figure*}

\begin{table*}
\centering
\caption{Pearson correlation coefficients and $p$-values for galaxy properties vs. \dSFRMS and $\rm \Delta f_{\him}$}
\label{t_pearson}
\begin{tabular}{lllllllllllllllllll}
\hline\hline
 & & \multicolumn{17}{c}{$\rm log\ (M_*/M_{\odot})$}\\
\cline{3-19} 
Fig. & y & \multicolumn{2}{c}{[9.0-9.5]}  && \multicolumn{2}{c}{[9.5, 10.0]}  && \multicolumn{2}{c}{[10.0, 10.5]}  && \multicolumn{2}{c}{[10.5, 11.0]}  && \multicolumn{2}{c}{[11.0, 11.5]}  && \multicolumn{2}{c}{Total}\\
\cline{3-4} \cline{6-7} \cline{9-10} \cline{12-13} \cline{15-16}  \cline{18-19}
 & & $c^a$  & $p$-value$^b$ && $c$  & $p$-value && $c$  & $p$-value && $c$  & $p$-value && $c$  & $p$-value  && $c$  & $p$-value \\
\hline
\multicolumn{19}{c}{x = \dSFRMS}\\
(3) & $f_{disk}$  & -0.09 & 2.40e-01 && 0.26  & 5.51e-04 && 0.22  & 2.35e-05 && 0.30  & 5.87e-08 && 0.30  & 6.15e-05 && 0.38  & 4.84e-41 \\
	  & $f_{bulge}$ & 0.08  & 2.62e-01 && -0.31 & 5.89e-05 && -0.27 & 2.64e-07 && -0.36 & 1.33e-10 && -0.36 & 2.28e-06 && -0.40 & 1.30e-47 \\
	  & $f_{bar}$   & 0.02  & 8.29e-01 && 0.05  & 5.01e-01 && 0.05  & 3.43e-01 && 0.07  & 2.20e-01 && 0.19  & 1.57e-02 && 0.04  & 1.96e-01 \\
(4) & $n_{\mathrm{S\acute{e}rsic}}$ & -0.20 & 9.22e-03 && -0.43 & 5.64e-09 && -0.55 & 1.67e-30 && -0.56 & 1.91e-26 && -0.40 & 6.14e-08 && -0.61 & 2.70e-122 \\
(5,6) & ($g-r$)$_{1kpc}$& -0.75 & 3.56e-33 && -0.51 & 2.79e-12 && -0.25 & 2.05e-06 && 0.02  & 6.67e-01 && 0.07  & 3.77e-01 && -0.48 & 3.65e-70 \\
        & ($g-r$)$_{1R_e}$& -0.78 & 8.74e-38 && -0.58 & 1.24e-16 && -0.44 & 1.24e-18 && -0.24 & 1.65e-05 && -0.21 & 6.04e-03 && -0.61 & 5.77e-119\\
      & ($g-r$)$_{2R_e}$& -0.79 & 2.17e-38 && -0.66 & 4.22e-22 && -0.56 & 1.51e-31 && -0.07 & 2.54e-01 && -0.49 & 1.11e-11 && -0.48 & 3.28e-68 \\
	  & ($g-r$)$_{1R_{e,ring}}$& -0.77 & 9.19e-36 && -0.63 & 1.51e-19 && -0.51 & 2.33e-25 && -0.39 & 9.49e-13 && -0.42 & 1.10e-08 && -0.66 & 9.61e-146 \\
	  & ($g-r$)$_{2R_{e,ring}}$& -0.65 & 1.41e-22 && -0.64 & 2.90e-20 && -0.57 & 3.13e-32 && -0.55 & 2.60e-25 && -0.60 & 1.04e-17 && -0.69 & 2.07e-167 \\
(7) & $\nabla(g-r)_{Re}$  & -0.04 & 5.67e-01 && -0.23 & 2.45e-03 && -0.32 & 5.77e-10 && -0.34 & 1.10e-09 && -0.38 & 4.28e-07 && -0.20 & 1.02e-11 \\
	  & $\nabla(g-r)_{2Re}$ & 0.13  & 9.46e-02 && -0.10 & 2.00e-01 && -0.13 & 1.07e-02 && -0.34 & 1.75e-09 && -0.43 & 6.49e-09 && -0.04 & 1.52e-01 \\
\hline
\multicolumn{19}{c}{x = $\rm \Delta f_{\him}^c$}\\
(8) & $f_{disk}$  & 0.15  & 5.02e-02 && -0.16 & 4.31e-02 && -0.14 & 8.54e-03 && -0.29 & 1.67e-07 && -0.50 & 6.47e-12 && -0.19 & 1.11e-10 \\
	& $f_{bulge}$ & -0.09 & 2.13e-01 && 0.23  & 2.52e-03 && 0.13  & 1.10e-02 && 0.31  & 2.29e-08 && 0.54  & 3.65e-14 && 0.22 & 3.09e-14 \\
	& $f_{bar}$   & -0.11 & 1.65e-01 && -0.12 & 1.33e-01 && 0.03  & 5.87e-01 && -0.01 & 7.99e-01 && -0.23 & 2.91e-03 && -0.05 & 5.93e-02 \\
(9) & $n_{\mathrm{S\acute{e}rsic}}$ & 0.25 & 8.41e-04 && 0.46 & 4.05e-10 && 0.49 & 2.31e-23 && 0.36 & 5.50e-11 && 0.30 & 6.10e-05 && 0.40 & 3.38e-46 \\
(10,11)  & ($g-r$)$_{1kpc}$& 0.51 & 4.08e-13 && 0.31 & 4.93e-05 && 0.22 & 1.82e-05 && -0.02 & 6.82e-01 && -0.06 & 4.05e-01 && 0.23 & 5.16e-16 \\
  & ($g-r$)$_{1R_e}$& 0.61 & 2.09e-19 && 0.47 & 2.29e-10 && 0.44 & 2.30e-18 && 0.17 & 3.49e-03 && 0.12 & 1.35e-01 && 0.37 & 4.30e-40 \\
  & ($g-r$)$_{2R_e}$& 0.75 & 4.49e-33 && 0.63 & 1.60e-19 && 0.60 & 2.99e-37 && 0.04 & 5.17e-01 && 0.41 & 3.31e-08 && 0.34 & 8.15e-33 \\
  & ($g-r$)$_{1R_{e,ring}}$& 0.72 & 5.58e-29 && 0.59 & 1.02e-16 && 0.52 & 5.12e-26 && 0.30 & 6.37e-08 && 0.32 & 2.58e-05 && 0.46 & 1.02e-62 \\
  & ($g-r$)$_{2R_{e,ring}}$& 0.80 & 7.67e-41 && 0.72 & 9.79e-28 && 0.70 & 8.80e-54 && 0.60 & 2.41e-31 && 0.55 & 7.31e-15 && 0.62 & 6.33e-127 \\
(12) & $\nabla(g-r)_{Re}$ & 0.36 & 1.08e-06 && 0.41 & 4.15e-08 && 0.36 & 2.07e-12 && 0.32 & 1.78e-08 && 0.42 & 1.32e-08 && 0.33 & 4.75e-32 \\
  & $\nabla(g-r)_{2Re}$ & 0.28 & 1.35e-04 && 0.39 & 2.55e-07 && 0.37 & 2.47e-13 && 0.57 & 5.77e-27 && 0.50 & 3.46e-12 && 0.35 & 4.57e-36 \\
\hline
\end{tabular}
\begin{flushleft}
Notes. $-$-- $^{a}$ Pearson correlation coefficients.
$^{b}$ Pearson $p$-value.
$^{c}$ \hi non-detections are included.
\end{flushleft}
\end{table*}


\subsubsection{Color Properties versus star formation}
In Figure \ref{fig5}, we use the $g-r$ color as an indicator of stellar population properties to examine its distribution across varying $M_*$ and $\rm \Delta SFR_{MS}$. 
For each galaxy, we measure the $g-r$ color using three characteristic radii: the central 1 kpc, $R_e$ and $2R_e$, where $R_e$ represents the $r$-band half-light radius.
The properties within the central 1 kpc of galaxies are commonly used to quantify bulge growth and to explore central stellar populations, which are closely linked to quenching processes \citep{Fang2013, Whitaker2017, Wang2020}. The half-light radius, $R_e$ (also known as the effective radius), represents the characteristic scale of galactic disks, while  $2R_e$ traces the outer regions of galaxies \citep[e.g.,][]{Shibuya2015, Miller2023}. In the low mass galaxies ($M_* \le 10^{10} M_{\odot}$), the typical size of  $R_e$ is $5.5 \pm 2.9$ arcsec, about 1.8 times the radius of 1 kpc. In the high mass galaxies ($M_* > 10^{10} M_{\odot}$), the typical size of $R_e$ is $4.3 \pm 1.9$ arcsec, roughly 3.3 times the radius of 1 kpc.

As a result, colors of each galaxy are measured within five specific regions: the central region with radius $R < 1\ \rm kpc$ ($(g-r)_{1kpc}$); an inner annulus from $1\ \rm kpc$ to $1R_e$ ($(g-r)_{1R_{e,ring}}$); an outer annulus from $1 R_e$ to $2 R_e$ ($(g-r)_{2R_{e,ring}}$); the entire inner region within $R < 1R_e$ ($(g-r)_{1R_e}$); and the extended region within $R < 2R_e$ ($(g-r)_{2R_e}$). 

Figures \ref{fig5} and \ref{fig6} illustrate color variations across different galactic regions, stellar mass bins, and levels of star formation activity. 
The colors across various galactic regions exhibit a complex relationship with stellar mass and the offset from the SFMS, in agreement with findings from prior studies \citep{Dimauro2022, Liao2023, Stephenson2024}. Generally, the outer regions of galaxies in our sample are bluer than their centers, and within the same mass range galaxies become bluer in most regions as SFR increases. However, low- and high-mass galaxies display distinct color distribution patterns. Specifically, in low-mass galaxies ($M_* \le 10^{10} M_{\odot}$), both the global and regional colors show strong negative correlations with $\rm \Delta SFR_{MS}$, with Pearson correlation coefficient of $c \sim$ -0.7 and $p$-value on the order of $\rm 10^{-20}-10^{-30}$. In high-mass galaxies ($M_* \ge 10^{10.5} M_{\odot}$), these trends become shallower, particularly in the central regions, where color remains nearly flat ($c \sim$ 0.0 and $p$-value $>$ 0.1). Additionally, color contrasts between the outer and central regions of galaxies tend to increase with stellar mass.

As an additional test, Figure \ref{fig6} compares the distribution of galaxies in the color-$\rm \Delta SFR_{MS}$ plane across different galactic regions. The correlation between $g-r$ color and $\rm \Delta SFR_{MS}$ shows a decreasing slope from lower to higher stellar masses and from outer to inner regions. The slope deviations between galaxies with varying stellar masses are more pronounced in the inner regions.

The difference in color between inner and outer regions is further illustrated in Figure \ref{fig7}, which shows the distributions of radial color gradients as a function of $\rm \Delta SFR_{MS}$. The color gradient is defined as the slope of the $(g-r)$ color against the normalized radius $\rm R/R_{e}$: $\rm \nabla(g-r) = \delta(g - r)/\delta(R/R_{e})$, where $\rm R_{e}$ is the effective radius in the $\it r$ band. We calculate two indices, $\rm \nabla(g-r){re}$ and $\rm \nabla(g-r){2re}$, over different radial ranges: $0.5\ \rm R_{e} < R < R_{e}$ and $\rm R_{e} < R < 2R_{e}$, respectively, to characterize the color gradients in the inner and outer regions of galaxies.

By examining Figure \ref{fig7}, we observe that the inner color gradients overall exhibit a similar distribution to the outer ones but are slightly lower (steeper). Specifically, the average inner color gradient is -0.09 $\pm$ 0.10, while the outer color gradient is -0.04 $\pm$ 0.06, indicating a modest but noticeable difference.
Most galaxies in our sample have negative color gradients, with low-mass galaxies showing less negative (shallower) gradients compared to high-mass galaxies. The difference is modest, ranging from $\sim$ 0.04 to 0.08. This result aligns with previous studies and supports the inside-out galaxy formation scenario \citep{Parikh2021, Liao2023}. However, due to the limited number of galaxies with stellar masses below $\rm 10^9 M_{\odot}$, we cannot reliably assess the color gradients at the low-mass end, where different behaviors might emerge \citep[e.g.,][]{Wang2022}. Additionally, an evident trend in Figure \ref{fig7} is that, for a given stellar mass, the color gradients become steeper with increasing $\rm \Delta SFR_{MS}$, but show a turning point around $\rm \Delta SFR_{MS} \sim 0$. For galaxies above the SFMS, i.e., starburst galaxies, the gradients become shallower. Specifically, the average slope is -0.08 for $\rm \Delta SFR_{MS} < 0$, steepens to -0.12 at $\rm \Delta SFR_{MS} \sim 0$, and then becomes shallower at -0.05 for $\rm \Delta SFR_{MS} > 0$. Consequently, the steepest color gradients are observed near this inflection point, particularly for galaxies with low to median stellar mass ($\rm M_* \sim 10^{10.5}M_{\odot}$).

\begin{figure*}
    \centering
    \includegraphics[width=1.0\hsize]{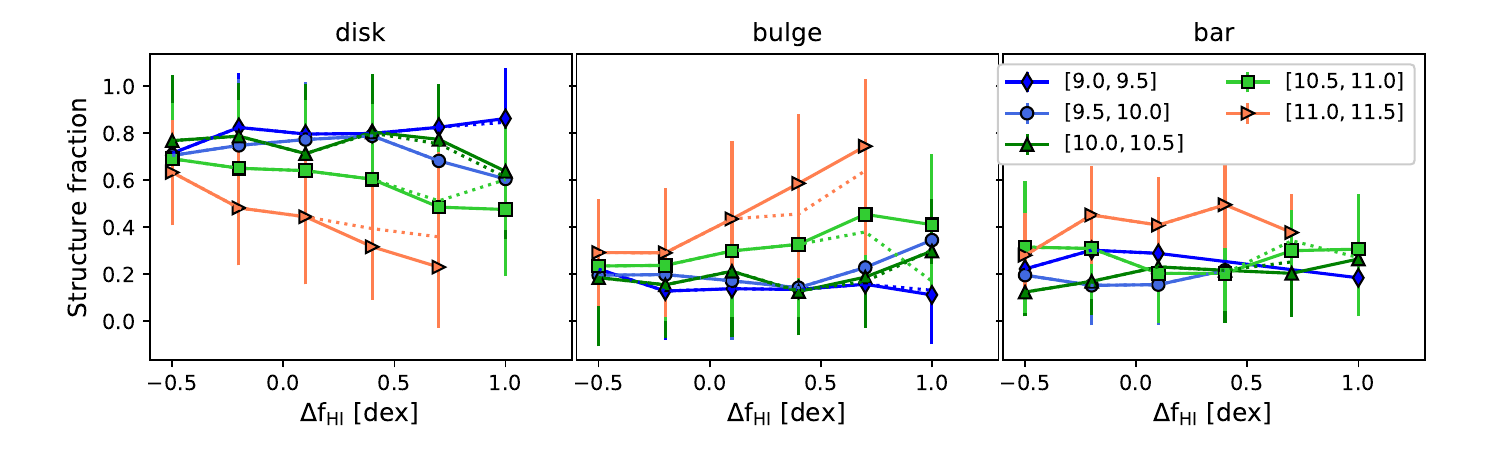}
	\caption{The relationship between structure flux fractions and \hi deficiency $\rm \Delta f_{\him}$. The stellar mass bins and symbols are the same as Figure \ref{fig3}. The dotted lines represent samples that include only \hi-detected galaxies.
	\label{fig8}}
\end{figure*}

\begin{figure}
    \centering
    \includegraphics[width=\hsize]{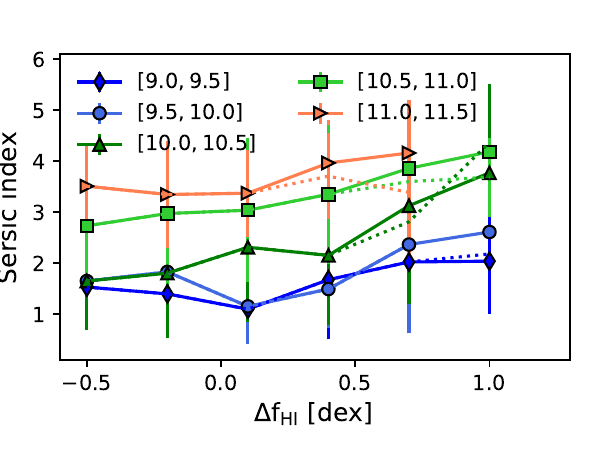}
	\caption{The relationship between global S\'{e}rsic index $\it n$ and \hi deficiency $\rm \Delta f_{\him}$. The stellar mass bins and symbols are the same as Figure \ref{fig4}. The dotted lines show samples with only \hi-detected galaxies.
		\label{fig9}}
\end{figure}

\begin{figure*}
    \centering
    \includegraphics[width=0.9\hsize]{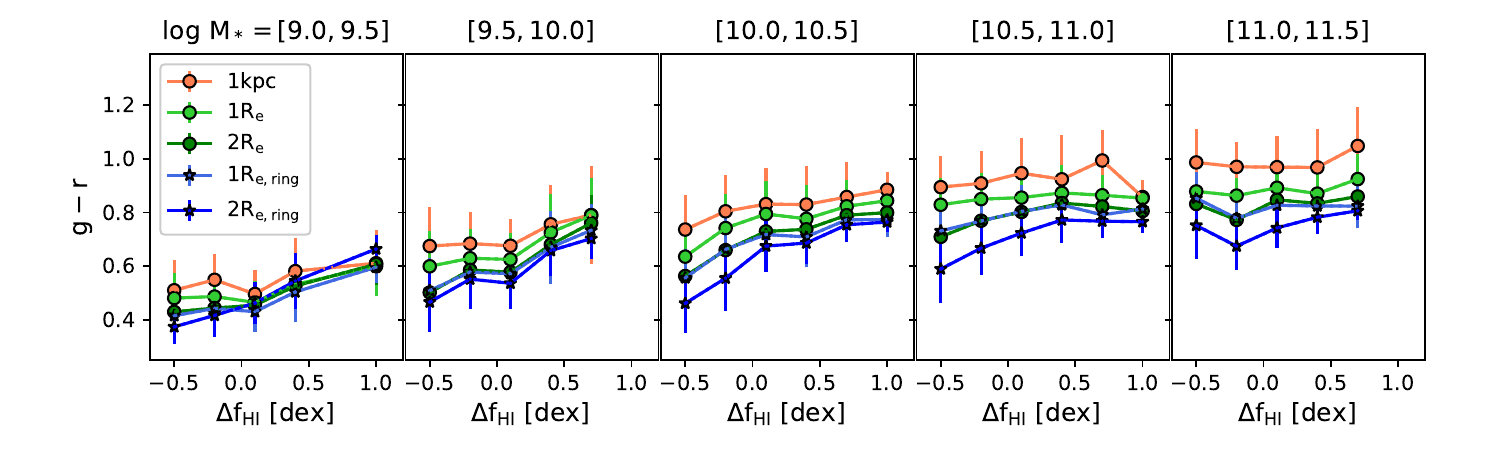}
	\caption{The relationship between galaxy color $g-r$ and \hi deficiency $\rm \Delta f_{\him}$. The stellar mass bins, color indexes, and symbols are the same as Figure \ref{fig5}. The dotted lines show samples with only \hi-detected galaxies.
		\label{fig10}}
\end{figure*}

\begin{figure*}
    \centering
    \includegraphics[width=0.9\hsize]{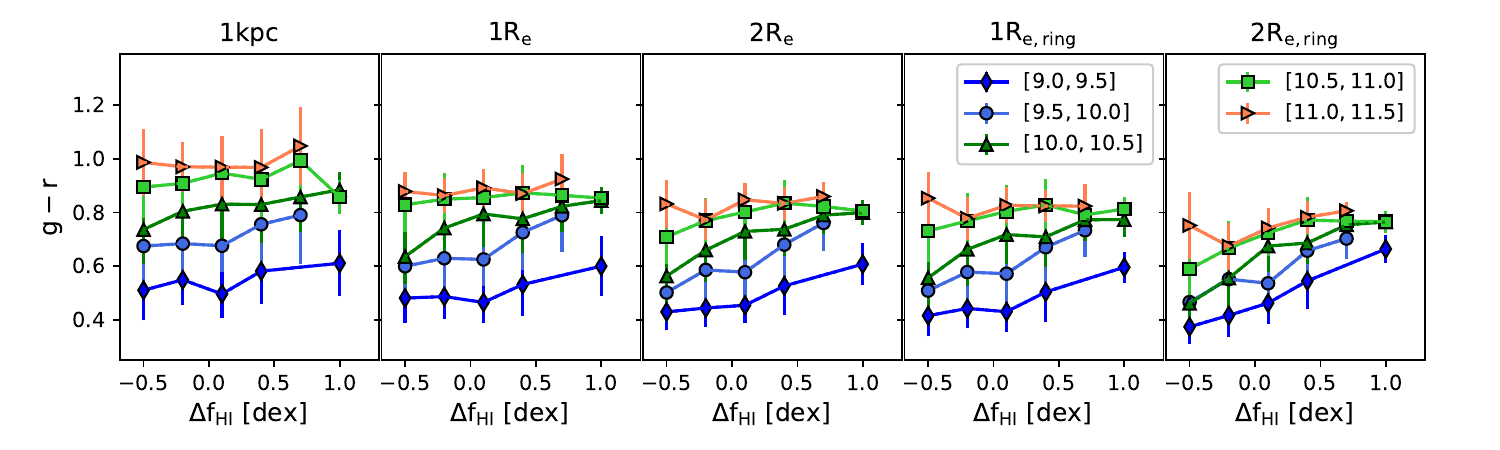}
	\caption{The relationship between galaxy color $g-r$ and $\rm \Delta f_{\him}$. The stellar mass bins and symbols are the same as Figure \ref{fig10}, but the colors from the same region of different stellar mass bins are put in a same panel. The symbols for stellar bins are shown in the legends of the right two panels.
		\label{fig11}}
\end{figure*}

\begin{figure*}
    \centering
    \includegraphics[width=\hsize]{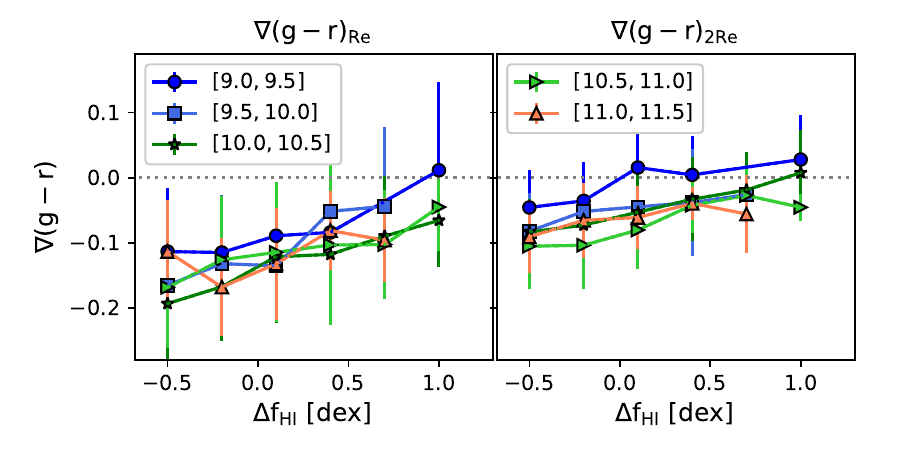}
	\caption{The relationship between radial color gradient and \hi deficiency $\rm \Delta f_{\him}$. The stellar mass bins, color gradients, and symbols are the same as Figure \ref{fig7}. The dotted lines show samples with only \hi-detected galaxies.
		\label{fig12}}
\end{figure*}

\subsection{Correlations between \hi gas and galaxy properties}
In the next step, we investigate the correlations between \hi gas content and structural properties of galaxies. It is well established that gas content plays a crucial role in the evolution of galaxies. Numerous studies have reported correlations between \hi gas fraction and various galaxy properties, such as stellar mass, color, bulge fraction, and stellar mass surface density \citep[e.g.,][]{Catinella2010, Chen2020}. To gain a clearer understanding of these correlations, we quantify \hi deficiency, which is defined as the deviation from the expected \hi gas fraction for a galaxy of a given stellar mass:
\begin{equation}
\rm \Delta f_{\him} = \langle log(M_{\him}/M_*)\rangle-log(M_{\him}/M_*).
\end{equation}
Here $\rm \log(M_{\him}/M_*)$ represents the observed \hi mass fraction, while $\rm \langle \log(M_{\him}/M_*) \rangle$ denotes the expected \hi fraction, derived from the \him-to-stellar mass scaling relation provided by \citet{Janowiecki2020}:
\begin{equation}
\rm \langle log(M_{\him}/M_*)\rangle = -0.53 \times log(M_*/M_{\odot}) + 4.7.
\end{equation}
In this context, galaxies with lower $\rm \Delta f_{\him}$ are considered \him-rich. For galaxies without \hi detection in our sample, the upper limit of \hi deficiency is estimated based on their observational detection limits.

\subsubsection{Structural Properties versus \hi gas}
In Figure \ref{fig8}, we present the bulge, disk, and bar fractions across five stellar mass intervals as a function of $\rm \Delta f_{\him}$. The dotted lines represent samples that include only \him-detected galaxies, while the solid lines include both detections and non-detections. The median values differ slightly depending on sample selection. The inclusion of \hi non-detections slightly shifts the median bulge fraction upward $\sim$ 10\%, particularly in \him-poor massive galaxies ($M_* > 10^{10.5} M_{\odot}$ and $\rm \Delta f_{\him} > 0.5$).

Statistical analysis shows that the bulge fraction exhibits a weak positive correlation with $\rm \Delta f_{\him}$ (Pearson correlation $c$ = 0.22, $p$ = 3.09E-14), while the disk fraction shows a weak negative correlation ($c$ = -0.19, $p$ = 1.11E-10). The bar fraction remains nearly constant, as indicated by its negligible correlation ($c$ = 0.08, $p$ = 0.17).
When \hi non-detections are excluded, the correlations become even weaker ($c$ = 0.08, $p$ = 0.031 for bulge fraction; $c$ = -0.08, $p$ = 0.033 for disk fraction; $c$ = 0.09, $p$ = 0.16 for bar fraction). 
Therefore, while the inclusion of \hi non-detections affects the statistical significance, the strength of the correlation remains weak overall.

However, for galaxies with stellar masses below $10^{10.5} M_{\odot}$, the correlation between bulge fraction and $\rm \Delta f_{\him}$ is weak (Pearson correlation $c$ = -0.1 - 0.1, $p$ $>$ 0.01 for each stellar mass bin), indicating that structural components remain nearly constant with increasing \(\hi\) deficiency. In contrast, for the most massive galaxies ($M_* \gtrsim 10^{10.5} M_{\odot}$), a significant positive correlation is observed ($c$ = 0.3-0.5 , $p$ $<$ 1e-8), suggesting that bulge fraction increases as \(\hi\) content decreases. 
Additionally, we examined the distributions of bulge fractions in two stellar mass regimes: $M_* > 10^{10.5} M_{\odot}$ and $M_* < 10^{10.5} M_{\odot}$. A Kolmogorov-Smirnov (K-S) test confirms that these distributions are significantly different (K-S statistic = 0.32, $p$-value = 4.57e-27). This mass $M_* \sim 10^{10.5} M_{\odot}$ is also consistent with the characteristic transition stellar mass seen in previous studies \citep[e.g.,][]{Kauffmann2003, Guo2008, Liao2023}.

This suggests that the relationship between \(\hi\) gas and bulge fraction likely differs between low-mass and high-mass galaxies. In low-mass galaxies, there appears to be little or no clear correlation between bulge fraction and \(\hi\) content at fixed stellar mass. In massive galaxies, a statistical correlation is observed, while this does not imply a direct causal link.
Other factors, such as environment, star formation history, or gas dynamics, may also play a role \citep[e.g.,][]{Oh2020, Sachdeva2020, Dimauro2022}. However, the limited resolution of \(\hi\) observations ($3^{\prime}$ beam size) prevents us from resolving the gas distribution within individual galaxies. As a result, we cannot directly associate \(\hi\) content with specific internal structures, such as bulges, disks, or bars. Thus, the observed correlation reflects global galaxy properties rather than detailed spatial relationships within galaxies.

Figure \ref{fig9} offers an additional check by using the global S\'{e}rsic index {\it n} as an indicator of galactic brightness profiles and structural characteristics. Across all stellar mass bins, the S\'{e}rsic index {\it n} shows a slight increase as galaxies become more gas-poor. This trend is consistent across both low and high stellar masses as a function of $\rm \Delta f_{\him}$. The median S\'{e}rsic index ranges from 1.3 in low-mass galaxies to 4.0 in high-mass ones. However, within each stellar mass bin, the variation remains small, with changes of less than 1 unit from gas-rich to gas-poor galaxies. 
Statistical analysis confirms that this correlation is weak. For the sample excluding \hi non-detections, the Pearson correlation coefficient $c$ is 0.25 with $p$-value  $<$ 0.001, indicating a weak positive trend. When including \hi non-detections, the correlation is slightly stronger ($c$ = 0.40, $p <$ 0.001), but the overall increase in the S\'{e}rsic index remains modest.  These results suggest that while more gas-poor galaxies likely tend to have higher S\'{e}rsic indices, the effect appears to be limited.

\subsubsection{Color Properties versus \hi gas}

Figures \ref{fig10} and \ref{fig11} illustrate the correlation between \hi gas content and the colors of galaxies in both inner and outer regions. We use the same markings and stellar mass bins as in Figures \ref{fig5} and \ref{fig6}. The distributions of galaxies in the color-$\rm \Delta f_{\him}$ plane resemble those in the color-$\rm \Delta SFR_{MS}$ plane, but the relationship is inverted and the trends are weaker. 

As expected, galaxies generally become redder in their inner and outer regions as they become more gas-poor (i.e., with increasing $\rm \Delta f_{\him}$). However, statistical analysis indicates that the strength of this trend varies across stellar masses. For the full sample, the Pearson correlation coefficient indicates a weak negative correlation between $\rm \Delta f_{\him}$ and color in the inner regions ($c$ = 0.23, $p$-value $<$ 0.001 ) and a moderate negative correlation in the outer regions ($c$ = 0.62, $p$-value $<$ 0.001 ). This correlation weakens in more massive galaxies, where increased scatter reduces the strength and significance of the relationship.

Additionally, Figure \ref{fig11} shows that the color of the outer disk region of galaxies exhibits a stronger dependence on \hi deficiency or richness compared to the inner region at a given stellar mass. The statistical analysis in Table \ref{t_pearson} confirms this trend, with a stronger Pearson correlation between $\rm \Delta f_{\him}$ and outer disk color than in the inner region.
At a fixed $\rm \Delta f_{\him}$, the scatter in color is also larger in the inner regions across different stellar mass bins. For instance, the $g-r$ color variation within a 1 kpc physical radius can be as large as $\sim$ 0.5 from the low to high stellar mass bins at a given $\rm \Delta f_{\him}$, whereas this variation reduces to $\sim$ 0.2 at a radius of 2$R_e$.

To investigate the relationship between \hi content and galaxy color, we conducted an analysis similar to that in Figure \ref{fig7}, but focused on examining the relationship between color gradient $\rm \nabla(g-r)$ and \hi deficiency $\rm \Delta f_{\him}$. Figure \ref{fig12} reveals a moderate correlation between the color gradient and \hi deficiency, with the Pearson correlation coefficient $c$ of 0.33-0.35 and $p$-value $<$ 0.001. 
Additionally, the median inner color gradient $\rm \nabla(g-r)_{Re}$ is -0.09 $\pm$ 0.10 mag/$R_e$, while the outer gradient $\rm \nabla(g-r)_{2Re}$ is -0.04 $\pm$ 0.06 mag/$R_e$. Although the inner regions appear slightly steeper on average, the difference is not statistically significant due to overlapping uncertainties.

For the entire galaxy sample, galaxies across different stellar mass bins exhibit similar slopes ($c \sim$ 0.33) and scatter ($\sigma \sim$ 0.1) in their color gradients (Figure \ref{fig12}, left panel). However, in \hi-poor galaxies ($\rm \Delta f_{\him} > 0$), we observe a systematic flattening in the $\rm \Delta f_{\him}$-$\rm \nabla(g-r)_{2re}$ relation, with the best-fit slope decreasing from $\sim$ 0.3 in \hi-rich galaxies to $\sim$ 0.2 in \hi-poor galaxies. This trend is evident across both low- and high-mass bins, suggesting that outer disk color gradients become less pronounced as galaxies lose their \hi reservoirs, potentially due to suppressed star formation in the outskirts.

\section{Discussion}
\label{sec:discuss}

Estimating structural properties is inherently challenging owing to various uncertainties. In this study, we use structural decompositions to analyze the properties of xGASS galaxies. Although alternative datasets have been developed to estimate these properties using SDSS images \citep{Simard2011, Meert2015, Cook2019}, our analysis leverages much deeper images from the DESI Legacy Imaging Surveys. These deeper images allow for more precise detection and fitting of stellar bars, improving the reliability of the structural decomposition. As a result, we achieve a more robust characterization of galaxy morphology, enabling a clearer assessment of correlations between structural properties, star formation activity, and \hi gas content. For instance, Figure \ref{fig4} shows that, despite the larger scatter, the distribution of bulge fraction in our sample follows a similar trend to that reported by \citet{Morselli2017}, even though our study is based on a significantly smaller number of galaxies.

Galaxy morphology plays a crucial role in shaping galaxy evolution. Numerous studies have explored the relationship between morphology, star formation, and cold gas reservoirs, with a particular focus on how bulge components influence star formation quenching \citep[e.g.,][]{Wuyts2011, Kauffmann2012, Buta2015}. Both observational and simulation studies suggest a strong link between quiescence and the presence of bulge structures \citep{Bluck2014, Snyder2015, Barro2017}. \citet{Dimauro2022} found that multi-band bulge–disk decompositions of massive galaxies reveal a clear link between bulge presence and the bending of the SFMS at high masses, suggesting that bulge development may be a key factor in galaxy quenching.



This evidence is also illustrated in Figures \ref{fig3} and \ref{fig4}, where the bulge fraction or S\'{e}rsic index $\it n$ generally increases from galaxies on the star formation main sequence (\dSFRMS $\sim$ 0) to those nearing quiescence (\dSFRMS $\sim$ -2) at a fixed stellar mass.
Pearson correlation analysis in Table \ref{t_pearson} confirms this trend, with the bulge fraction and S\'{e}rsic index negatively correlated with \dSFRMS ($c$ = -0.40, $p$-value $<$ 0.001 for bulge fraction; $c$ = -0.61, $p$-value $<$ 0.001 for S\'{e}rsic index).

Additionally, Figure \ref{fig5} shows that most galaxies have redder centers compared to their outskirts, with more massive galaxies appearing redder overall. Since color is often used as a proxy for  star formation activity and stellar population characteristics \citep{Gonzalez-Perez2011}, this trend may reflect a higher prevalence of older stellar populations or earlier quenching of star formation in central regions compared to disks \citep{Ibarra-Medel2016, Jegatheesan2024}.

Furthermore, a comparison between Figures \ref{fig3} and \ref{fig7} shows that galaxies on the SFMS tend to have steeper negative color gradients and lower bulge fractions (i.e., higher disk fractions). This pattern is consistent with the inside-out formation scenario, where gas gradually accretes onto the galaxy outskirts, sustaining star formation over time \citep{Wang2011, Kennedy2016, Morselli2017, Liao2023}. However, this alignment does not serve as direct evidence of a specific formation pathway.

Figure \ref{fig3} further indicates that the relationship between bulge fraction and \dSFRMS is not strictly monotonic. A subset of galaxies on the upper envelope of the MS have bulge fractions of approximately 30\%, which are higher than those of galaxies along the MS \citep{Morselli2017}. 
As shown in Figures \ref{fig4} and \ref{fig7}, galaxies with \dSFRMS $>$ 0 tend to have bulges with relatively low S\'{e}rsic indices ($\it n$ $\sim$ 1-2) and shallower color gradients ($\rm \nabla(g-r)$ $\sim$ -0.05). Their color gradients are less steep than those of galaxies on the SFMS ($\rm \nabla(g-r)$ $\sim$ -0.12).
This observation suggests that star formation may also be occurring within these bulges, although it predominantly takes place in the outskirts \citep{Breda2020, Stephenson2024}. 
A similar trend is found in  \citet{Liao2023}, where galaxies with high specific SFRs tend to exhibit shallower or even positive color gradients, which may indicate extended star formation histories.

The interplay between \hi content and galaxy morphology is crucial for understanding morphological transformation and star formation quenching. Previous studies have demonstrated that gas reservoirs play a key role in shaping galaxy structure and star formation activity \citep[e.g.,][]{Zhang2009, Catinella2010}. To further explore this relationship, we examine correlations between morphological properties and \hi gas content.

Figures \ref{fig8} and \ref{fig9} show that when \hi non-detections are included, gas-poor galaxies tend to have higher bulge fractions and S\'{e}rsic indices at a fixed stellar mass. Statistical analysis supports this trend, with Pearson $p$-values both less than 0.001. This result aligns with the expectation that \hi gas is predominantly associated with actively star-forming, disk-dominated systems \citep{Catinella2015, Catinella2018}. However, when the analysis is limited to \hi detections only, the correlation coefficients become small (e.g., $c \sim 0.08$), indicating that the correlations are weak despite their formal statistical significance. This suggests that the observed trends are not strongly driven by galaxies with \hi detections alone and may also be sensitive to other factors.

Furthermore, Figure \ref{fig11} shows that the $g-r$ color within a 1 kpc radius exhibits significant scatter ($\sigma \sim$ 0.5), with a nearly flat slope (Pearson $c$ $\sim$ 0.23) with respect to $\rm \Delta f_{\him}$. In contrast, the outer disk color exhibits a stronger dependence on \hi fraction (Pearson $c \sim$ 0.62 and $p$-value $<$ 0.001), suggesting that \hi content is more linked to the outer disk than to the bulge \citep{Fabello2011, Cook2019, Chen2020}.

This relationship is further supported by Figure \ref{fig12}, which shows a clear positive correlation between color gradients and $\rm \Delta f_{\him}$ (Pearson $c \sim$ 0.33 and $p$-value $<$ 0.001). As galaxies become more gas-poor, the median color gradient flattens from $-0.15$ to $-0.05$ mag/$R_e$. These results are consistent with previous studies showing that \hi-rich galaxies tend to have bluer, actively star-forming outer disks compared to their inner regions \citep{Wang2011, Chen2020}. This trend likely reflects an inside-out formation scenario, where \hi gas is accreted at large radii, fueling star formation in the outer disk over time \citep{Sancisi2008, Pilkington2012, Padave2024}.

Despite the advantages of using deeper images, some limitations of this work should be acknowledged. First, the increased sensitivity may introduce additional challenges in distinguishing faint structures from background noise, potentially affecting the uncertainties in bulge–disk decomposition and bar identification. Second, the sample selection, based on xGASS galaxies, may not fully represent the morphological diversity of the broader galaxy population. Additionally, the resolution of the \hi observations in our work is still relatively low, making it difficult to spatially associate \hi content with specific internal structures.

Future studies incorporating larger datasets and alternative decomposition methods could help address these limitations.
Upcoming wide-field surveys such as LSST \citep{Ivezic2019}, Euclid \citep{Euclid2022}, Roman \citep{Akeson2019}, and CSST \citep{Zhan2011, Gong2019} will provide deeper optical and infrared imaging. Additionally, next-generation \hi surveys, including FAST \citep{Nan2011, Jiang2019} and WALLABY \citealt{Koribalski2020}, will offer higher sensitivity for detecting neutral hydrogen. These advancements will enable more comprehensive studies of the interplay between galaxy structure, star formation, and cold gas content across a wider range of environments and redshifts.

\section{Summary}
\label{sec:summary}

In this paper, we utilize multi-band images from the DESI Legacy Imaging Surveys to perform structural decompositions of xGASS galaxies. After masking and cleaning signals from unassociated objects, such as foreground stars and background galaxies, we conduct isophotal analysis to derive the surface brightness profiles of the galaxies, identify stellar bars, and perform detailed 2D multi-component decomposition. We thoroughly investigate how the structural properties correlate with star formation activity and \hi gas content, as well as the connection between galaxy colors and these properties. Our primary results are summarized as follows.

\begin{itemize}
\item The bulge fraction generally decreases with increasing \dSFRMS\ and in lower-mass galaxies, indicating that star-forming galaxies are predominantly disc-dominated, while quiescent systems are bulge-dominated. However, the bar-to-total fraction shows no significant variation across different stellar mass bins or \dSFRMS. Additionally, we observe a slight increase in the bulge-to-total ratio for galaxies with $\rm \Delta SFR_{MS} > 0$, suggesting that these galaxies are not purely disk-dominated but instead contain a population with intermediate bulge fractions.

\item The correlation between color $g-r$ and \dSFRMS\ shows a decreasing slope from low to high stellar mass and from outer to inner regions, along with larger color variations. This trend suggests that younger stellar populations are generally concentrated in the disks of low-mass galaxies undergoing active star formation.

\item The galaxies in our sample generally exhibit negative color gradients, with shallower gradients observed in less massive galaxies and in the outer regions. Additionally, there is an inflection point in the relationship between color gradient and \dSFRMS, where color gradients become steeper from passive to star-forming galaxies, reaching their lowest values at around \dSFRMS\ $\sim$ 0 in any stellar mass bin, and then becoming shallower above the SFMS.

\item The bulge fraction of galaxies increases slightly with higher \hi deficiency, while the disk fraction decreases, mirroring the trend of the S\'{e}rsic index {\it n}. However, this trend is less pronounced in low-mass galaxies, where the proportions of various structural components remain almost unchanged with increasing \hi deficiency at a fixed stellar mass.

\item At a fixed stellar mass, galaxies generally become redder as they become more gas-poor, although there is a large scatter among massive galaxies. Additionally, the color of the galactic outer disk region shows a stronger dependence on \hi fraction than the inner region.

\item We find a clear correlation between color gradient and $\rm \Delta f_{\him}$, where the color gradients become flatter with increasing $\rm \Delta f_{\him}$ (i.e., as galaxies become more \hi-poor). Additionally, gas-poor galaxies tend to have flatter color gradients in their outer disk regions.

\end{itemize}
\section*{Acknowledgements}
\label{sec:acknow}
We thank the anonymous referee and editor for their critical comments and instructive suggestions that significantly improved the quality of the paper.
This work was supported by the National Key Research and Development Program of China (No. 2022YFA1602902), the China Manned Space Project (Nos. CMS-CSST-2025-A08, CMS-CSST-2025-A19), and the Chinese National Natural Science Foundation grant (No. 12073035).

The DESI Legacy Imaging Surveys consist of three individual and complementary projects: the Dark Energy Camera Legacy Survey (DECaLS), the Beijing-Arizona Sky Survey (BASS), and the Mayall z-band Legacy Survey (MzLS). DECaLS, BASS and MzLS together include data obtained, respectively, at the Blanco telescope, Cerro Tololo Inter-American Observatory, NSF’s NOIRLab; the Bok telescope, Steward Observatory, University of Arizona; and the Mayall telescope, Kitt Peak National Observatory, NOIRLab. NOIRLab is operated by the Association of Universities for Research in Astronomy (AURA) under a cooperative agreement with the National Science Foundation. Pipeline processing and analyses of the data were supported by NOIRLab and the Lawrence Berkeley National Laboratory (LBNL). Legacy Surveys also uses data products from the Near-Earth Object Wide-field Infrared Survey Explorer (NEOWISE), a project of the Jet Propulsion Laboratory/California Institute of Technology, funded by the National Aeronautics and Space Administration. Legacy Surveys was supported by: the Director, Office of Science, Office of High Energy Physics of the U.S. Department of Energy; the National Energy Research Scientific Computing Center, a DOE Office of Science User Facility; the U.S. National Science Foundation, Division of Astronomical Sciences; the National Astronomical Observatories of China, the Chinese Academy of Sciences and the Chinese National Natural Science Foundation. LBNL is managed by the Regents of the University of California under contract to the U.S. Department of Energy. The complete acknowledgments can be found at https://www.legacysurvey.org/acknowledgment/.

\software{Astropy \citep{Astropy2013}, Matplotlib \citep{Hunter2007}, NumPy \citep{Harris2020}, SciPy \citep{Virtanen2020}, SExtractor \citep{Bertin1996}, IRAF \citep{Tody1986, Tody1993}, GALFIT \citep{Peng2010}, GALFITM \citep{Vika2013, Haussler2022}}

\bibliography{main}{}
\bibliographystyle{aasjournal}

\end{document}